\documentclass[12pt, a4paper]{article}
\usepackage[utf8x]{inputenc}
\usepackage{longtable}
\usepackage[square,sort,comma,numbers]{natbib}

\usepackage{amsmath,amsfonts, amsthm, amssymb}
\usepackage{newlfont}
\usepackage{fancyhdr}
\usepackage[Glenn]{fncychap}
\usepackage{fullpage}

\usepackage[T1]{fontenc}
\usepackage{array,multicol}
\usepackage[pdftex]{graphicx}
\usepackage{appendix}
\usepackage[author={jo}]{pdfcomment}

\usepackage{setspace} 
\usepackage{xcolor}
\usepackage[colorinlistoftodos]{todonotes} 

\DeclareOldFontCommand{\rm}{\normalfont\rmfamily}{\mathrm}
\DeclareOldFontCommand{\sf}{\normalfont\sffamily}{\mathsf}
\DeclareOldFontCommand{\tt}{\normalfont\ttfamily}{\mathtt}
\DeclareOldFontCommand{\bf}{\normalfont\bfseries}{\mathbf}
\DeclareOldFontCommand{\it}{\normalfont\itshape}{\mathit}
\DeclareOldFontCommand{\sl}{\normalfont\slshape}{\@nomath\sl}
\DeclareOldFontCommand{\sc}{\normalfont\scshape}{\@nomath\sc}

\def\R{\mat{R}}

\def\Kk{{\cal K}}

\def\DD{{\cal D}}
\newcommand{\Db}{{\boldsymbol D}}

\def\K{{\cal K}}

\newcommand{\nnabla}{{\boldsymbol \nabla}}
\newcommand{\Fb}{{\boldsymbol F}}
\newcommand{\Lb}{{\boldsymbol L}}

\newcommand{\bs}{{\boldsymbol \sigma}}
\newcommand{\bbs}{{\boldsymbol b}}

\newcommand{\bm}{{\boldsymbol m}}
\newcommand{\bu}{{\boldsymbol v}}
\newcommand{\D}{{\boldsymbol D}}
\newcommand{\W}{{\boldsymbol W}}
\newcommand{\bW}{\boldsymbol  W}

\newcommand{\Rb}{{\boldsymbol R}}
\newcommand{\M}{{\boldsymbol M}}
\newcommand{\Pb}{{\boldsymbol P}}

\newcommand{\Sb}{{\boldsymbol S}}
\newcommand{\Ub}{{\boldsymbol U}}

\newcommand{\Ib}{{\boldsymbol I}}

\newcommand{\Qb}{{\boldsymbol Q}}
\newcommand{\bV}{{\boldsymbol V}}

\newcommand{\I}{{\boldsymbol I}}

\newcommand{\be}{{\boldsymbol e}}

\newcommand{\bn}{{\boldsymbol n}}
\newcommand{\bc}{{\boldsymbol c}}

\newcommand{\bD}{{\boldsymbol D}}
\newcommand{\ff}{{\boldsymbol f}}

\def\R{\mathbb{R}}
\newcommand\bfdiv{\mathop{\bf div}\nolimits}

\newcommand\sign{\mathop{\rm sign}\nolimits}
\numberwithin{equation}{section}

\usepackage{hyperref}
\hypersetup{
	colorlinks   = true,    
	urlcolor     = blue,    
	linkcolor    = blue,    
	citecolor    = blue     
}

\begin{document}
	
	\title{Large plastic deformation of voids in crystals}
	\author{Jalal Smiri\footnote{LSPM,   University Sorbonne-Paris-Nord, Villetaneuse, France} \footnote{Universite Marie et Louis Pasteur, CNRS, Institut FEMTO-ST, F-25000 Besan\c{c}on, France},  Joseph Paux$^*$, O\u{g}uz Umut Salman$^*$ and Ioan R. Ionescu$^{*}$\footnote{IMAR, Romanian Academy, Bucharest, Romania}
	}
	\date{version : \today}
	\maketitle
	
	
	\begin{abstract}
		
		
		The mechanisms of void growth and coalescence are key contributors to the ductile failure of crystalline materials. At the grain scale, single crystal plastic anisotropy induces large strain localization leading to complex shape evolutions. In this study, an Arbitrary Lagrangian-Eulerian (ALE) framework for 2D crystal plasticity combined with dynamic remeshing is used to study the 2D shape evolution of cylindrical voids in single crystals. The large deformation and shape evolution of the voids under two types of loading are considered: (i) radial and (ii) uni-axial loadings. In both cases, the voids undergo complex shape evolutions induced by the interactions between slip bands, lattice rotations and large strain phenomena. 
		
		In  case (i), the onset of the deformation revealed the formation of a complex fractal network of slip bands around the voids. Then, large deformations unearth an unexpected evolution of the slip bands network associated with significant lattice rotations, leading to a final hexagonal shape for the void. In case (ii), we obtain shear bands with very large accumulated plastic strain  ($>200\%$) compared to the macroscopic engineering strains ($<15\%$). A high dependence between crystalline orientations, slip band localization and therefore shape evolution was observed, concluding in a high dependency between crystalline orientation and void shape elongation, which is of prime importance regarding coalescence of the voids, thus to the formation of macro-cracks.

		\noindent {\textbf{keywords} : Void growth, Crystal plasticity, Slip bands, Single crystal, large deformation, Eulerian approach}    
	\end{abstract}
	
	\newpage 

	\newpage
	
	\section {Introduction} 
	The mechanisms of void growth and coalescence are key contributors to the ductile failure of crystalline materials, occurring through plastic flow around pre-existing voids or nucleated cavities at second-phase particles. These cavities can emerge on two distinct scales within polycrystalline material: at the grain scale (i.e., intragranular and intergranular porosities) and at the polycrystalline scale. The progression of void growth is highly dependent on the nature of the surrounding plastic deformation, which can be modeled by (i) crystal plasticity at the grain scale and (ii) isotropic or texture-induced anisotropic plasticity at the polycrystal scale. 
	Thus, micromechanical modeling of void growth and coalescence must account for these varying local behaviors in order to adequately capture the influence of porosity on the overall macroscopic response. However, it is generally infeasible to derive comprehensive analytical models for the plasticity of ductile porous solids due to the complexity involved.

	The aim of this  paper is to carry out an analysis of  large deformation  voids growth  in single crystals (HCP, FCC)  using  Eulerian crystal plasticity finite element simulations. This analysis is driven by previous work on the analytical homogenization of porous single crystals with internal cavities, as demonstrated in several models that address void growth \cite{kysar2005cylindrical,HAN20132115, PAUX20151, paux2022, MBIAKOP2015436, LING201658, SONG2017560, SRIVASTAVA201767, KHAVASAD2021103950}. In particular, it extends the work of Paux et al. (2022), who contribute to the understanding of ductile failure in crystalline materials by addressing the anisotropic plastic behavior of single crystals using an innovative approach that includes a new class of piecewise constant velocity fields derived from numerical FFT analysis restricted to small strain, which effectively improves the estimation of the macroscopic yield stress, particularly in models with periodic void distributions in single crystals.
	
	While the model proposed in \cite{paux2022} provides a robust framework to estimate yield criteria, it lacks the ability to predict the evolution of void shapes during deformation, a critical factor in understanding the long-term behavior of materials under stress. To address this shortcoming, the present paper  makes use of  an Eulerian  numerical model that not only captures the onset of yielding but also accurately predicts the evolution of void shapes.
	We aim to extend the contributions of \cite{paux2022} by providing a more comprehensive tool for the analysis of ductile failure in crystalline materials. The model is applied to the well-known experimental study by \cite{crepin1996}, which observed that voids in an HCP crystal evolve into a polyhedral cavity, a prediction also made in the same study. In that work, void growth was successfully modeled using a finite element approach that incorporated elastic-plastic behavior and kinematic hardening, with growth rates associated with prismatic slip activation. In this context, the proposed Eulerian rigid (visco) plastic proposed model offers an alternative or complementary approach for analyzing void evolution.

	Conventionally, large deformation simulations are performed in a Lagrangian framework, where the mesh follows the deforming material. While effective for moderate strains, as the deformation progresses, mesh distortion introduces significant errors and ultimately leads to simulation failure. To address these challenges, this paper adopts an Eulerian approach, which is particularly effective for simulating extreme deformations. In the Eulerian formulation, a fixed grid allows materials to flow through without distorting the mesh, making it ideal for analyzing complex material behavior during severe plastic deformation.
	
	
	The elastic-(visco-)plastic model  needs a permanent  interplay between both Eulerian (more appropriate  for plasticity) and Lagrangian (more appropriate for elasticity)  descriptions. However,  in applications involving large deformations of metals, the elastic component of the deformation is small  with respect to the inelastic one,  and can be neglected  by using  a rigid-(visco)plastic approach (see for instance  \cite{hut76,lebtome93,kok02}).  This simplified model takes important theoretical and numerical  advantages  by using only one  (Eulerian) configuration.  This work starts from
	the  Eulerian framework in modeling large deformations in crystal
	plasticity introduced by \cite{CI09} and uses the high-resolution
	numerical methods developed in \cite{Cazacu2010-hi}.  
	


	\bigskip

	Let us outline the contents of the paper. The aim of the  first section is to recall from \cite{CI09} the 3D rigid-viscoplastic  crystal model   and its 2D simplification. In the next two sections, the proposed model is evaluated under two distinct plane-strain configurations involving intergranular porous crystals : (1) Growth evolution under radial loading (hydrostatic tension) applied on a hexagonal close-packed (HCP) crystal and (2) Growth evolution under uniaxial loading (tensile test) applied to a face-centered cubic (FCC) crystal. These cases are chosen to illustrate the model's capability in different crystallographic structures and loading conditions, thereby demonstrating its broader applicability. Finally, after the conclusion, an  appendix  recalls  the numerical algorithm form  \cite{Cazacu2010-hi} and  details the re-meshing procedure used to obtain the results.

	\section{ Eulerian rigid-plastic approach of  crystal  plasticity} \label{Chapter:model}

	Plastic deformation in crystalline materials is highly complex and involves multiple length scales. These spatial heterogeneities range from the atomistic scale (dislocation cores and grain boundary structures) to mesoscale dislocation patterns and grain microstructures, extending to the macroscopic scale of the specimen. At the macroscopic level, plasticity manifests itself as a smooth flow described by a continuous stress-strain response.
	
	Several theories have been developed to describe and understand plastic deformation at the macroscopic level. Continuum Crystal Plasticity (CP) theory is the most widely used approach for modeling crystal plasticity. This theory incorporates lattice-based kinematics into the classical continuum framework. Its original mathematical formulation was introduced by Hill \cite{Hill1966-xr} and Hill and Rice \cite{Hill1972-vl}, with initial applications by Asaro and Rice \cite{Asaro1977-nn,Asaro1983-cw} and Pierce, Asaro, and Needleman \cite{Peirce1983-sm}. Since then, many authors have further developed CP theory (see \cite{Roters2010-fv} for a comprehensive overview). The classical Continuum crystal theory of plasticity assumes that crystalline materials undergo irreversible flow when applied stresses exceed certain thresholds. This theory models the stress-strain response with continuous curves, implying that the  heterogeneities are averaged out. This approach has been successful in reproducing key plasticity phenomena such as yielding, hardening, and shakedown.

	CP theory spans macroscopic length scales, from micrometers ($\mu$m) to millimeters (mm), due to its coarser representation of plastic deformation compared to other methods such as molecular dynamics or discrete dislocation dynamics approaches. In recent CP formulations, deformations are considered finite, and the continuum description clearly distinguishes between reference and deformed configurations.

	\subsection{Multiplicative decomposition of the deformation gradient}
	
	Consider a single crystal at time $t=0$,  free of any surface tractions and body forces and let choose this configuration, say  $\Kk_0 \subset \R^d$ (here $d=2,3$ is the space dimension) as reference configuration of the crystal. Let $\Kk=\Kk(t)\subset \R^d$ denote the current configuration. The incorporation of lattice features is achieved through a multiplicative decomposition of the total deformation gradient $\Fb$ into elastic and plastic components:
	\begin{equation}
		\Fb = \Fb^{e} \Pb.
		\label{eq:1}
	\end{equation}
	This decomposition implies a two-stage deformation process. First, $ \Pb$ transforms the initial reference state $\Kk_0$ to an intermediate state $\tilde{\Kk}$, characterized by plastic deformation only with no change in volume.   $\Pb$ is called the (visco)plastic deformation with respect to the reference configuration of a material neighborhood of the material point $X$ at time $t$. Then, $\Fb^{e}$ brings the body to the final configuration $\K$ through elastic deformation and rigid lattice rotation, i.e. $\Fb^e=\Rb\Ub^e$ where $\Rb$ denotes the rotation of the crystal axes with respect to its isoclinic orientation, and $\Ub^e$ is the elastic right stretch tensor. 
	

	Following \cite{Mandel1972-gk,Asaro1977-nn}, $\Pb$ is assumed to leave the underlying lattice structure undeformed and unrotated, ensuring the uniqueness of the decomposition in \eqref{eq:1}. The unique feature of CP theory is its construction of the plastic component $\Pb$ by constraining dislocation kinematics. Plastic flow evolves along pre-selected slip directions via volume-preserving shears, leaving the crystal lattice undistorted and stress-free \cite{McHugh2004-kl}.

	Since in applications involving large deformations and high strain rates, the elastic component of the deformation is small  with respect to the inelastic one, it can be neglected and   a rigid-viscoplastic approach will be adopted (such a hypothesis is generally used e.g.  Hutchinson \cite{hut76},  Lebensohn and Tome \cite{lebtome93},  Kok et al.\cite{kok02}, etc. ). That means that  we  neglect  the elastic lattice strain $\Ub^e$ by supposing that  $\Ub^e \approx  \Ib $.  This leads to the following decomposition for the deformation gradient $\Fb$ (see for example Kok et al. \cite{kok02}): 
	\begin{equation}
		\Fb=\Rb\Pb.
		\label{RP}
	\end{equation}
	Such a hypothesis is valid since during forming or other industrial processes, the elastic component of deformation is negligibly small (typically 10$^{-3}$) in comparison to the plastic component (typically $>$10$^{-1}$). It is also to be noted that once the elastic/plastic transition is over, the stress evolution in the grains is controlled by plastic relaxation (see  Tom\'e  and Lebensohn \cite{lebtome93}).

	\subsection{Eulerian description  of  the lattice rotations}
	
	Crystal slip systems are labeled by integers $s= 1,...,N$, with $N$ denoting  the number of slip systems. Each slip system $s$ is  specified by the unit vectors $(\bbs_s^0, \bm_s ^0)$, where $\bbs_s^0$ is the slip direction and $\bm_s^0$ is the normal to the slip plane in the perfect undeformed lattice.  Since the viscoplastic deformation does not produce distortion or rotation of the lattice, the  lattice orientation is the same in the reference and intermediate configurations $\Kk_0$ and $\tilde{\Kk}$ and is specified by  $(\bbs_s^0, \bm_s^0 ), s= 1...N$.

	Let's note   $\bbs_s=\bbs_s(t)$ and $\bm_s=\bm_s(t)$ the glide direction and glide plane normal, respectively, in the deformed configuration $\Kk$. At $t=0$, it reads $\bbs_s(0)=\bbs_s^0$  and $\bm_s(0)=\bm_s^0$ . Since elastic effects are neglected,
	\begin{equation}\label{evlnoi}
		\bbs_s = \Rb \bbs_s^0, \quad   \bm_s = \Rb \bm_s^0.
	\end{equation}
	Note that (\ref{evlnoi}) implies that $\bbs_s$ and   $\bm_s$ are unit vectors. Furthermore,
	\begin{equation}\label{prod}
		\bbs_s\otimes \bm_s = \Rb\left(\bbs_s^0 \otimes \bm_s^0\right)\Rb^T. 
	\end{equation}

	We seek  to express the lattice evolution equations only in terms of vector and tensor fields associated with the current configuration. Let  $\bu=\bu(t,x)$ be the Eulerian velocity field,  $\Lb$ the velocity gradient,  $\D$ the rate of deformation,  and $\W$ the spin tensor,
	\begin{equation}\label{DW}
		\Lb=\Lb(\bu)=\nabla \bu, \quad  \D=\D(\bu)=(\nabla \bu)^{symm}, \quad \W=\W(\bu) =(\nabla \bu)^{skew}.
	\end{equation}
	
	The viscoplastic deformation is only due to slip ; the slip contribution to the viscoplastic deformation being (\cite{ric71}, \cite{TeodSid76}),
	\begin{equation}\label{flowrulenoi}
		\dot{\Pb}\Pb^{-1} = \sum_{s=1}^N\dot{\gamma}^{s}\bbs_s^0\otimes\bm_s^0,
	\end{equation}
	where  $\dot{\gamma}^{s}=\dot{\gamma}^{s}(t)$ is the viscoplastic shear rate on the slip system $s$.    If we  denote by  
	\begin{equation}\label{MR}
		\M_s= \left(\bbs_s\otimes\bm_s\right)^{symm}, \quad \Qb_s=\left(\bbs_s\otimes\bm_s\right)^{skew}. 
	\end{equation}
	then,   using $\Lb= \dot{\Fb}\Fb^{-1}= \dot{\Rb}\;\Rb^{T} + \Rb\dot{\Pb}\Pb^{-1}\;\Rb^{T}$,  Eqs.  (\ref{prod})   and  (\ref{flowrulenoi}), the rate of deformation $\D$   can be  written as
	\begin{equation}\label{D}
		\D = \sum_{s=1}^N \dot{\gamma}^{s}\M_s. 
	\end{equation}
	Taking the  anti-symmetric  part of  $\Lb$,  we  obtain that the spin tensor is $\W=\dot{\Rb}\;\Rb^{T}+\sum_{s=1} ^N\dot{\gamma}^{s} \Qb_s$  and a differential equation for the rotation tensor $\Rb$:
	\begin{equation}\label{Rp}
		\dot{\Rb}=(\W-\sum_{s=1} ^N\dot{\gamma}^{s} \Qb_s)\Rb. 
	\end{equation} 
	The evolution equations (\ref{Rp})  describe the evolution of the lattice in terms of vector and tensor fields associated with the current configuration.

	\subsection{Plastic and  visco-plastic flow rules }
	\label{flowruled}
	
	In order to complete the model, we need to provide the constitutive equation for the slip rate $\dot{\gamma_s}$ as a function of $\tau_s$, the stress  component acting on the slip plane of normal 
	$\bm_s$ in the slip direction $\bbs_s$.
	In the current configuration, $\tau_s$ is expressed as
	\begin{equation}\label{eq:taus}
		\tau^s = \bs : \M_s, 
	\end{equation}
	where $\bs=\bs(t)$ is the Cauchy stress tensor acting in the current configuration $\K$ while $\M_s$ is defined by (\ref{MR}). Note that  $\{\tau^s\}_{s=\overline{1,N}}$ are not independent; they belong to a fifth dimensional space  of $\R^N$ corresponding to the dimension of the space  of deviatoric stresses. 
	
	To determine the shear strain rates $\dot{\gamma}_{\alpha}$ relative to the local stress, a constitutive law is needed. Various proposals exist, ranging from phenomenological to more physically based approaches. One simple phenomenological approach assumes that $\dot{\gamma}_{\alpha}$ depends on the stress only through the resolved shear stress $\tau^s$. 
	The associated plastic dissipation functional is neither strongly convex nor differentiable, so the solution could not be unique. Additional assumptions are needed in order to restrict the number of solutions.

	One way to overcome this difficulty of determining the active slip systems problem is to  adopt a rate-dependent approach for the constitutive response of the single crystal.
	A widely used  rate-dependent (viscoplastic) model is the Norton type model, which  relates the shear strain rate $\dot{\gamma}^{s}$ on a slip system $s$  to the resolved shear stress   $\tau^s $  through a power-law   (see Asaro and Needleman \cite{asaneed85}) 
	\begin{equation}\label{shear_strain1}
		\dot{\gamma}^{s}=\dot{\gamma}_{0}^{s}\,{\left\vert\frac{\tau^s}{\tau_c^s}\right\vert}^n\,\sign(\tau^s),
	\end{equation}
	where $\dot{\gamma}_{0}$ is a reference shear strain rate,  while the exponent $\emph{n}$ has a fixed value.    The Schmid model is recovered  for large values of $n$ ($n \to \infty$). 
	
	Another regularization  of the Schmid law, which will be used here in the numerical computations,   can be done by using a  Perzyna-like viscoplastic law  of the form:
	\begin{equation}\label{flow_rule}
		\dot{\gamma}^s = \dfrac{1}{\eta_s}\left[\vert \tau^s \vert-\tau_c^s \right]_+ \sign(\tau^s), 
	\end{equation}
	where $\eta_s$ is the  viscosity, which may depend on the slip rate,  and $[ \;  x\; ]_+=(x+|x|)/2\;$ denotes the positive part of any real number $x$.  The Schmid model is recovered  for  vanishing viscosities $\eta_s$ (i.e. $\eta_s \to 0$).

	\subsection{2-D model with 3 slip systems} \label{ToyModel}
	
	As a first approach to consider the shape evolution of voids in single crystal under large strain, a 2-D model ($d=2$) is considered. This simplified model (by comparison with 3-D models) provides the possibility to investigate closely the formation of slip bands and the evolution of the lattice rotation around the void with reasonable computation time, therefore enabling an interpretation of the link between crystal plasticity features and shape evolution of the voids.  
	
\begin{figure}[ht]
		\begin{center} 
			\includegraphics[height=7.cm, angle=0]{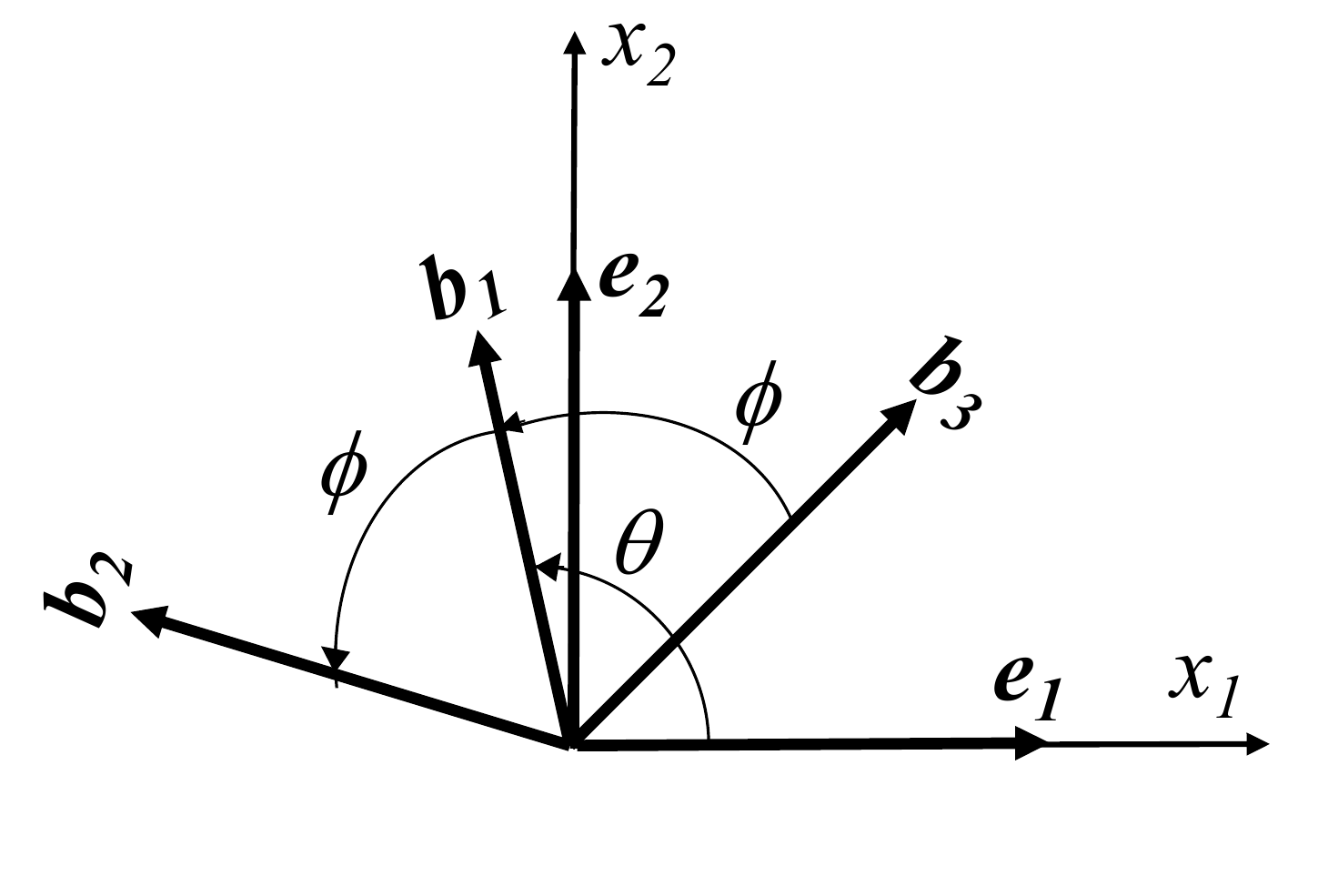}
		\end{center}
		\caption{Two dimensional model with three slip systems.} \label{3slip}
	\end{figure}

	We will  introduce here   a two dimensional  model  with $N=3$ slip systems.  
	In this case let us denote with $\phi$ and  $-\phi$   the angles  formed by slip system $r=1$ with the other two systems $r=2,3$ and let  $\theta$ be the  angle formed by the slip system $1$ with the $Ox_1$ axis (see Figure \ref{3slip}).   The three composite in-plane slip systems  $ \bbs_1, \bbs_2,  \bbs_3$ are  specified  by  the angles  $\theta,  \theta+\phi,  \theta-\phi$.
	\label{sec:TOYMODEL}
	
	\bigskip 
	
	The main simplification  for the 2-D problem comes from the  lattice rotation  $\Rb$ which is now a rotation  $\Rb(\theta, \be_3)$ with angle $\theta$ along $0x_3$ axis and  we have 
	\begin{equation}\label{Rb}
		\Qb_r=\dfrac{1}{2}\left((1,0)\otimes(0,1) -  (0,1)\otimes(1,0)\right).
	\end{equation}
	for all $r=1,2,3$. (\ref{Rp}) simplifies into a scalar form, 
	\begin{equation}\label{theta}
		\dot{\theta} = \frac{\partial \theta}{\partial t} + \bu\cdot \nabla \theta= \frac{1}{2}\left( \sum_{r=1}^{3}\dot{\gamma}^{r} -(\frac{ \partial v_1}{ \partial x_2}-
		\frac{ \partial v_2}{ \partial x_1})\right).
	\end{equation}
	
	\bigskip
	
	As already stated in the literature, this model is physically sound in  two situations : i) the in-plane  deformation of a FCC crystal \cite{gan2006cylindrical}, and ii) the slip in the basal plane of a hexagonal crystal \cite{gan2007cylindrical}. 
	
	\bigskip
	{\bf In-plane  deformation of a FCC crystals} 
	
	\bigskip
	Rice \cite{ric71} showed that certain pairs of the  three-dimensional systems that are potentially active need to combine in order to achieve plane-strain deformation.   For an FCC crystal, with 12 potentially active slip systems,    we consider  $Ox_3$ axis to be parallel to  $ [1 1 0] $ in the  crystal basis, which  means that   the plane-strain  plane $(Ox_1x_2)$ is   the plane  $[\bar{1}10]-[001]$.   Some geometrical constraints  (see \cite{Cazacu2010-hi})  have to are satisfied such that  $N=3$ pairs of   composite  systems will give deformation in the plane $(Ox_1x_2)$.   For all $r=1,2,3$ we have denoted by $\bbs_r$  and $\bm_r$ the normalized projections of the corresponding three-dimensional slip directions and normal directions $(k, l)$  onto the $x_1x_2$-plane. The angle $\phi$  between  the slip systems $1$ and $2$ is given by
	\begin{equation}\label{t123}
		\phi=\arctan(\sqrt{2}) \approx 54.7^\circ. 
	\end{equation}
	%
	
	Since there are  some scalar factors between the first two components  of the in-plane systems and the 3-dimensional ones given by  
	$	q_1=\frac{1}{\sqrt{3}}, \quad q_2=q_3= \frac{\sqrt{3}}{2},$
	the 2-D composite slipping rate $\dot{\gamma}^r$ corresponds to the  3-D  slipping rate $\dot{\gamma}^k$  multiplied by $2q_r$.  As it follows from \cite{Cazacu2010-hi},  the  2-D yield limit   $\tau_0^r$ corresponds to the  3-D yield limit   $\tau_0^k$  divided by $q_r$ while the  2-D   viscosity   $\eta^r$ corresponds to the  3-D yield limit   $\eta^k$  divided by $q_r^2$.

	\bigskip
	{\bf Slip in the basal plane of a hexagonal crystal}  
	\bigskip
	
	Alternatively, this situation corresponds to hexagonal close-packed (HCP) crystals (such as Ti, Mg, Zr, etc) under plane strain, with the plane of deformation aligned with the basal plane (0001) of the hexagonal lattice. This situation has been considered experimentally in \cite{crepin1996}. In this situation, the plastic strain is mainly accommodated by the three prismatic slip systems, i.e. the $(10\bar{1}0)\langle1\bar{2}10\rangle$ slip family. Each of those slip systems can act independently from the two others, as its strain is a plane strain. One might remark that combinations of other slip systems of HCP crystals from the basal and the pyramidal slip families could also lead to plane strain, but it would require significantly more energy, and therefore never occur. Thus, those slip families are not considered in this work. The prismatic slip systems are symmetric, leading to  
	\begin{equation}
		\phi=\pi/3= 60^\circ.
	\end{equation}


	\subsection{Initial and boundary value problem formulation}
	
	\label{BVP}

	We begin by presenting the equations governing the  motion in a domain $\DD=\DD(t)$  of an incompressible rigid-viscoplastic crystal for the simplified 2D model with $N=3$ slip systems.  In an Eulerian description of a crystal visco-plasticity theory, the unknowns    are: the velocity $\bu: [0,T]\times \DD \to \R^2$, the crystal lattice orientation, i.e. the rotation $\theta: [0,T]\times \R$ t  and the Cauchy stress $\bs : [0,T] \times \DD \to \R^{2\times2}_S$. Let   $\bs =\bs'+p\I$, where $\bs'$ is the stress deviator while $p : [0,T] \times \DD \to \R$ is the pressure.

	The momentum balance  in the Eulerian coordinates reads
	\begin{equation} \label{mbl}
		\rho^{mass}(\partial_t \bu + \bu \cdot \nnabla \bu)  - \bfdiv \bs'  + \nabla p= \rho^{mass}{\ff} \quad
		\hbox{ in } \DD,
	\end{equation}
	where   the mass density $\rho^{mass} >0$ and the body forces $\ff$ are supposed to be known. The  incompressibility condition is
	\begin{equation}\label{inc}
		\mbox{div} \,\bu =0 \quad \hbox{ in } \DD.
	\end{equation}
	The momentum balance equations are completed by the  constitutive equation,  which relates the stress tensor $\bs$ and the rate of deformation tensor $\D(\bu)$ (see (\ref{D})) through the evolution equations for each slip system $s$  given by the viscous regularizations (\ref{shear_strain1}) or (\ref{flow_rule}).
	
	The boundary $\partial \DD$ of the domain  $\DD$  is decomposed into two disjoint parts, $\Gamma_v$ and $\Gamma_s$, such that the velocity is prescribed on     $\Gamma_v$ and traction is prescribed   on  $\Gamma_s$, at any time $t$:
	\begin{equation} \label{bcv}
		\bu(t) = {\bf V}(t) \quad \hbox{on} \quad \Gamma_v, \quad \bs(t) \bn = \Sb(t)
		\quad \hbox{on} \quad
		\Gamma_s,
	\end{equation}
	where $\bn$ stands for the outward unit normal on $\partial \DD$,
	$\bV$  is the imposed velocity and $\Sb$ is the prescribed
	stress vector.

	We  also consider another partition of  $\partial \DD$
	into $\partial_{in} \DD(t)$ and $\partial_{out} \DD(t)$
	corresponding to  incoming ($\bu\cdot \bn <0$)  and outcoming  ($\bu\cdot \bn \geq 0$)  flux.  The boundary
	conditions associated to  the lattice  evolution  equations  (\ref{theta}) reads 
	\begin{equation}\label{bcbmTM}
		\theta(t) = \theta^{in}(t),   \quad     \quad \hbox{on} \quad
		\partial_{in} \DD(t),
	\end{equation}
	
	To  the field equations, we add the initial conditions
	\begin{equation}\label{ic}
		\bu(0)=\bu^0, \quad \theta(0) = \theta^0,  \quad   \hbox{in} \quad
		\DD.
	\end{equation}
	where $\bu^0$ is the initial velocity and  $\theta^0$ gives the initial orientation of  the crystal lattice.  Note that in this model,  there is no need to prescribe initial conditions for the stress. This is very convenient since the initial stress field is generally not known or it cannot be easily measured.  
	
	\subsection{Numerical strategy}
	
	To integrate the governing equations,  a  mixed finite-element and Galerkin discontinuous  strategy,  developed in  \cite{CI09}, will be used.   We will give here only the principles of the method used in the next sections while a  detailed description is provided in Appendix \ref{Ann:Numerical}. Time implicit (backward) Euler scheme  is used  for the field equations, which 
	gives  a set of nonlinear equations for the velocities $\bu$ and lattice orientation $\Rb$.  
	At each  iteration in time,  an iterative  algorithm is used in order to solve these nonlinear equations. Specifically, the variational  formulation  for the velocity field is discretized using the finite element method, while a  finite volume method with an upwind choice of the flux is adopted for solving the hyperbolic equations that describe the evolution of the lattice orientation. One of the advantages of the proposed numerical strategy is that it does not require the consideration of elastic deformation since it is not based on an elastic predictor/plastic corrector scheme (e.g. \cite{SH}).
	
	It is to be noted that in the case of the proposed rigid-viscoplastic model  (\ref{flow_rule}), numerical difficulties arise from the non-differentiability of the viscoplastic terms.  That means that one cannot  use  finite element techniques developed for Navier-Stokes fluids (see for instance \cite{P,tem}).  To overcome these difficulties the iterative  decomposition-coordination formulation coupled with the augmented Lagrangian method of  \cite{GlLT,FoGl}  was  modified. 
	The reason for this modification  is that in the used crystal model, there is not co-axiality between the stress deviator and the rate of deformation as it is the case in the Bingham model used by \cite{GlLT,FoGl}.  
	Note that this  iterative decomposition-coordination algorithm permits  to solve  at each iteration  the equations  for  the unit vectors that define the lattice orientation. 
	
	Furthermore, if the   domain $\DD$  occupied by the single crystal (or poly-crystals) varies in  time,  then an arbitrary Eulerian-Lagrangian  (ALE)  description was adopted.  In the numerical simulations presented in the next sections  we  use   the followings spatial  discretization: P2  for the velocity field, P1 for the pressure, P1 discontinuous for the stress field, slip rates, dislocation densities and lattice orientations.

	\section{Void growth evolution under radial loading} 
	
	\subsection{Setting of the numerical experiment}
	
	In the present section, we present the numerical setting, including geometrical and material characteristics, initial and boundary conditions, followed by the results and discussion of the 2D void growth simulations under plane strain in an HCP single crystal. 
	
	\begin{figure}
		\centering{}
		\includegraphics[height=8cm]{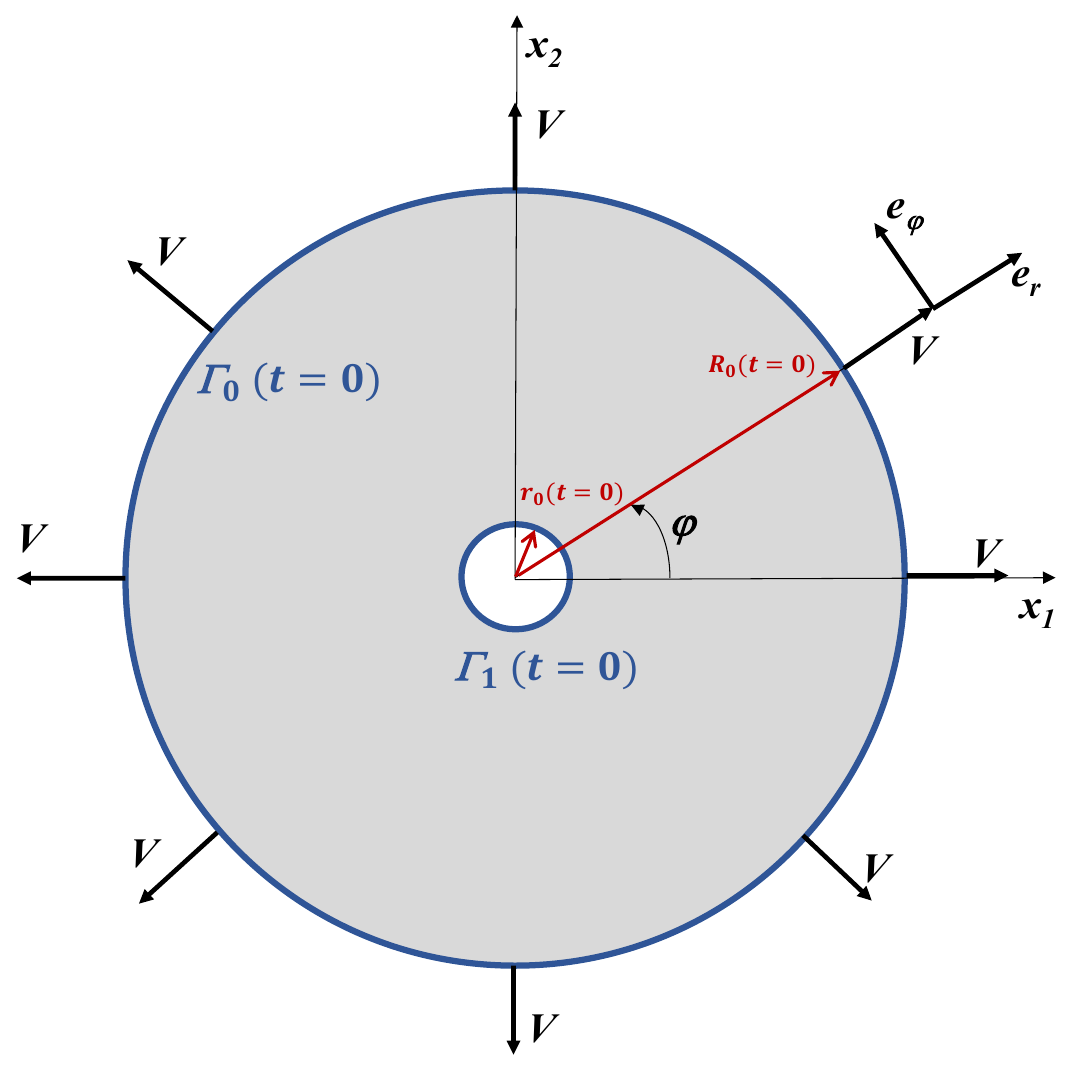}
		\caption{ Schematic representation (not to scale) of a porous single-crystal disc unit cell with a single circular void at its center.}
		\label{fig:shematiccell}
	\end{figure}

	Fig. \ref{fig:shematiccell} shows a schematic representation of the initial state of the one-void unit cell represented by a disc of radius $R_0$, containing an initially circular void at its center with a radius $r_0=R_0/30$. This configuration results in an initial void fraction of $\displaystyle f_{0} =\left(\frac{r_0}{R_0}\right)^{2}= 0.111 \%$. 

	On the external boundary, denoted by $\Gamma_0(t)$,  we impose an  expansion velocity $V_0$, corresponding to the in-compressible Eulerian  velocity field $\displaystyle \bu^*=\frac{V_0R_0}{r}\be_r$, i.e. 
	\begin{equation}
		{\bf V}(t)(x,y)=\frac{V_0R_0}{x^2 + y^2}(x\be_1+y\be_2),  \quad \mbox{for} \; (x,y)\in \Gamma_0(t), 
	\end{equation}
	while the internal boundary, denoted by $\Gamma_1(t)$, is traction free (i.e. $\Sb(t)=0$). 
	The initial conditions in velocity and in the  crystal orientation where chosen to be 
	\begin{equation}
		\bu^0 (x,y)= \frac{V_0R_0}{x^2 + y^2}(x\be_1+y\be_2), \quad \theta^0(x,y)=\theta_0.
	\end{equation}
	The expansion velocity $V_0$ was chosen to be   small such that the process is expected to be quasi-static.  By integration on the Eulerian coordinate $r$  gives  the external radius  can be deduced to be $R^2(t)=R_0^2+2V_0R_0t$, which means that  the engineering area deformation is given by 
	\begin{equation}
		\epsilon^{eng}(t)=\frac{R(t)^2-R_0^2}{R_0^2}=\frac{2V_0t}{R_0}.
	\end{equation}	
	The time interval $[0,T]$, the velocity  $V_0$ and the rayon $R_0$ where chosen such that the  final engineering volume/area deformation  is  $\epsilon^{eng}_{final}=1.32\%$.  Note that, even if the final $\epsilon^{eng}_{final}$ is small, the strains  around the cavity are  expected to be much larger.  Based on the engineering area deformation formula above, we estimate  it to be of order of  $\epsilon^{eng}_{final}R_0/r_0=30\epsilon^{eng}_{final}\approx  40\%$. The material parameters used in this quasi-static computation is not directly specific to a particular material but reflect the order of magnitude typically associated with hexagonal close-packed (HCP) crystals. The material parameters are: density $\rho = 3200\,\text{kg/m}^3$, slip resistance $\tau_c = 20\, \text{MPa}$ for all systems with no hardening, and a viscosity  $\eta$ chosen to be as small as possible (of order of  $1\%$  of $\tau_{c} R_0/V_0$) for numerical reasons. The initial orientation of the crystal was chosen to be $\theta_0=0^\circ$.
	
	In the following subsections, we examine the evolution of slip bands around the cavity, from their initiation to large deformations under hydrostatic tension loading as predicted by the proposed numerical model. Given the distinct phases observed in the deformation fields, the analysis is divided into two stages, each discussed in a separate subsection. The first stage focuses on small deformations, analyzing the onset of shear and kink bands and comparing the findings with existing models from the literature. The second stage explores the large deformation regime, following the evolution  of the deformation patterns,   described in the onset phase.

	\subsection{Onset of the shear/kink band deformation}
	
\begin{figure}[h]
		\centering{}
		\includegraphics[height=5.5cm]{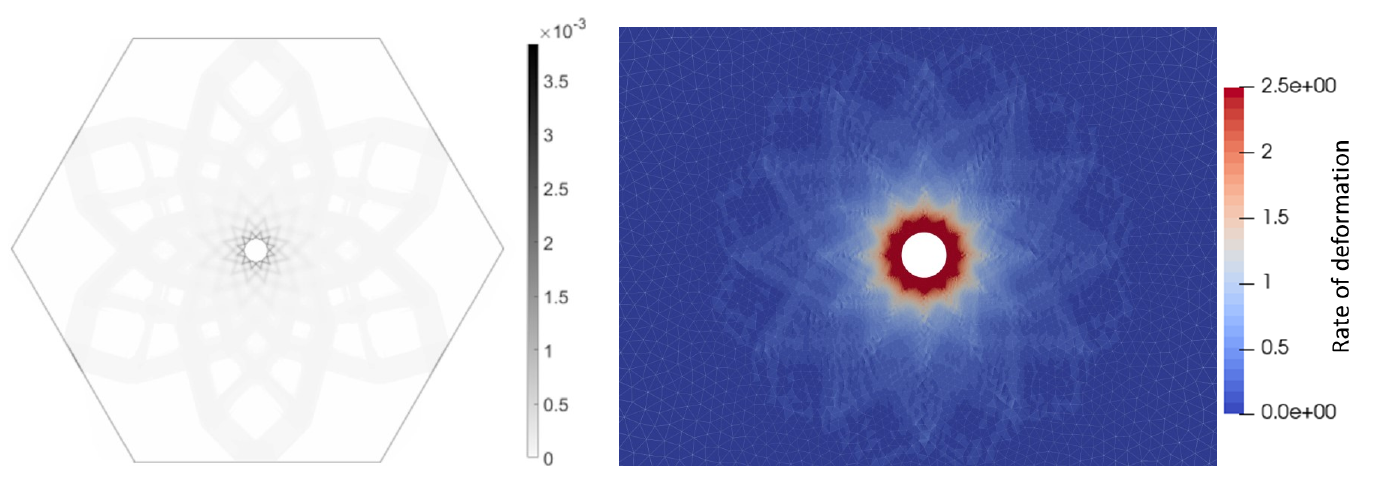}
		\caption{Onset of the deformation: qualitative comparison using two different approaches. Left: equivalent  strain obtained under small strain hypothesis, as reported in \cite{paux2022}. Right:  zoom of the equivalent strain rate at a  low deformation level ($\epsilon^{eng}_{vol}=0.0033\%$) obtained with the  Eulerian model.} 
		\label{fig:hex_defeq}
	\end{figure}

	At low deformation levels, although the size and morphology of the void show minor variations, slip lines are present, allowing for their characteristics to be analyzed. In Figure \ref{fig:hex_defeq}, we present a qualitative comparison of the equivalent von Mises strain distribution in a periodic hexagonal unit cell with aligned crystal orientation $\theta=0^\circ$, computed using the FFT method described in \cite{paux2022} with the deformation rate predicted by the current model. This comparison is particularly valid at low deformation levels where the two models are valid and where the localization patterns of deformation and deformation rate are similar. Despite the difference in cell micro-structure,   hexagonal in \cite{paux2022} and disk-shaped in the current model,  this comparison remains justified at equivalent low porosity levels (0.111$\%$ and 0.24$\%$, respectively), where edge effects are minimal and do not significantly impact the results around the cavity. All results are presented with an enlarged view around the void of the circular cell to clearly visualize the different fields. The zoomed region is shown as a rectangular area, with dimensions set to $26 r_{0}$ in width and $19 r_{0}$ in height, centered on the void, where $r_{0}$ is the initial radius of the void.

	At the onset of strain localization, a fractal network emerges as observed in the spatial distribution of strain and strain rate (Figure \ref{fig:hex_defeq}). A (redundant) base pattern of twenty four slip lines, composed of two superimposed six-pointed stars with a 45-degree phase shift between them, recursively repeats himself with self-similarity from the exterior of the cell up to the void, forming successive layers.	The base pattern is made up of two types of lines: lines parallel to slip directions, known as classical slip bands, and lines parallel to the slip plane normals, referred to as kink bands. 
	
	A slip band is defined as a region characterized by high plastic strain and low crystallographic reorientation, where a single slip system is predominantly active and the slip direction is parallel to the band. In contrast, a kink band is a region with high plastic strain and sharp crystallographic reorientation, also dominated by one slip system, but with the Burgers vector perpendicular to the band direction \cite{Forest1998-iw,zecevic2022new}.
	
	The consistent appearance of this twenty-four-slip-line pattern in both models suggests that it is an intrinsic feature of the strain field in porous single crystals. Notably, there is a clear agreement between the two models in their description of the emergence  of slip and kink bands.
	
\begin{figure}
		\begin{center}
			\includegraphics[height=7cm]{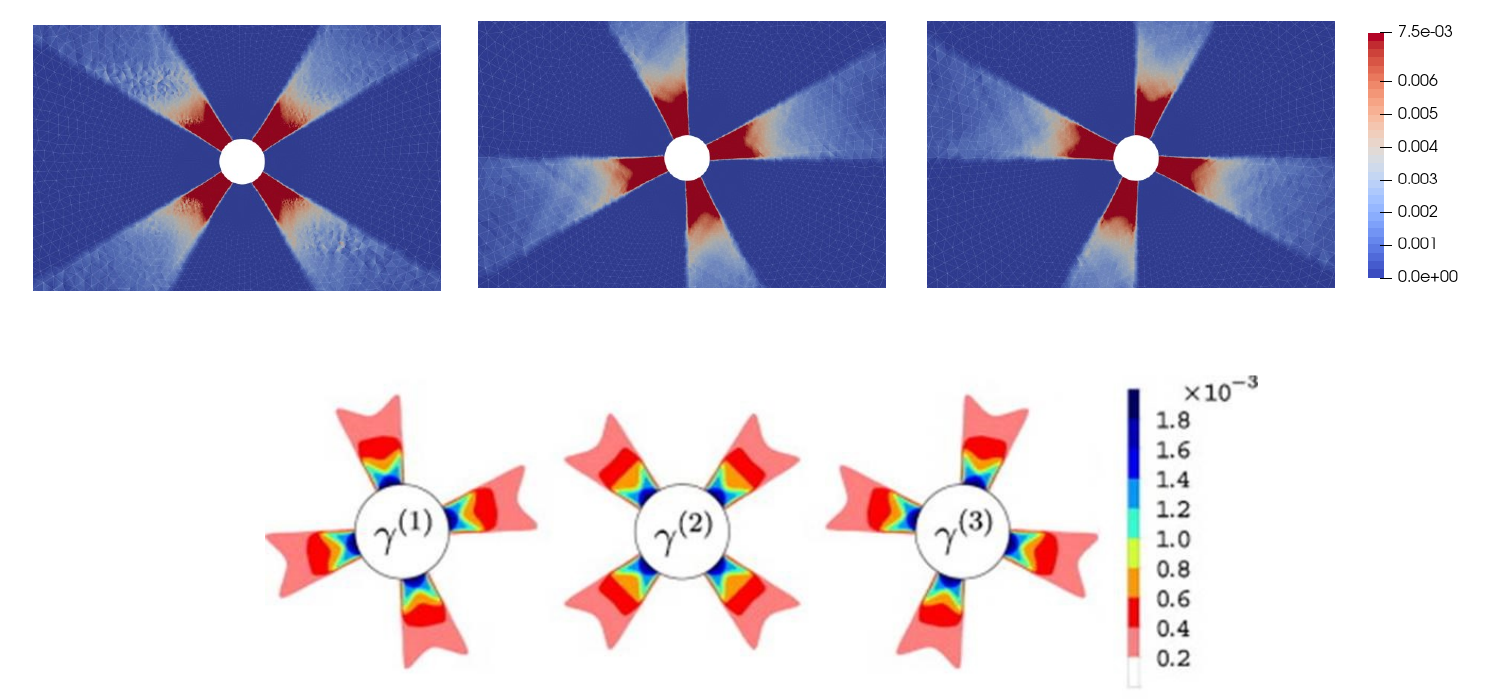}
			\caption{Onset of the deformation. The three slip systems of HCP crystal based on finite element simulation results : (left)  $s=1$, (middle) $s=2$, and (right)  $s=3$. Top:   slip deformation rates computed with the presented Eulerian model.  Bottom:   slips deformation  computed with a strain gradient crystal plasticity   model  without hardening  from \cite{BORG20076382}.  }
			\label{fig:hex_slip}
		\end{center}

	\end{figure}

	In this context of incipient plasticity, the three prismatic slip systems demonstrate equivalent amounts of glide, as shown in Fig. \ref{fig:hex_slip} (Top), with all slip systems being simultaneously active. 
	Each system is active in specific angular areas, forming, for each system, an orthogonal cross. The crosses of the different systems are equivalent considering the crystal symmetries.    The system's activities maxed out at the contact with the void. A similar finding   is  reported by \cite{BORG20076382} (see Figure \ref{fig:hex_slip} bottom), where it is noted that slip contours are periodic with a period of $90^\circ$. Specifically, slip system one is predominantly active in the region $0 \le \theta \le 30^\circ$, slip system two in $30 \le \theta \le 60^\circ$, and slip system three in $60 \le \theta \le 90^\circ$ of the upper right quadrant.

	\bigskip
	
	Having explored the initiation and early evolution of slip and kink bands at small deformation levels and seen an agreement between the present model and the model of  Paux et. al.  \cite{paux2022}, we now shift our focus to the large deformation regime. In this stage, the behavior of the slip bands undergoes significant changes as deformation progresses, with noticeable effects on the microstructure and deformation patterns. In the following section, the response of the model is used to examine how these patterns evolve under increasing strain, highlighting the transformation of slip lines and the eventual disappearance of shear bands and/or kink bands.
	
	\subsection{Large deformations  towards a hexagonal shape}
	
	\begin{figure}[h]
		\begin{center}
			\qquad
			\includegraphics[height=12cm]{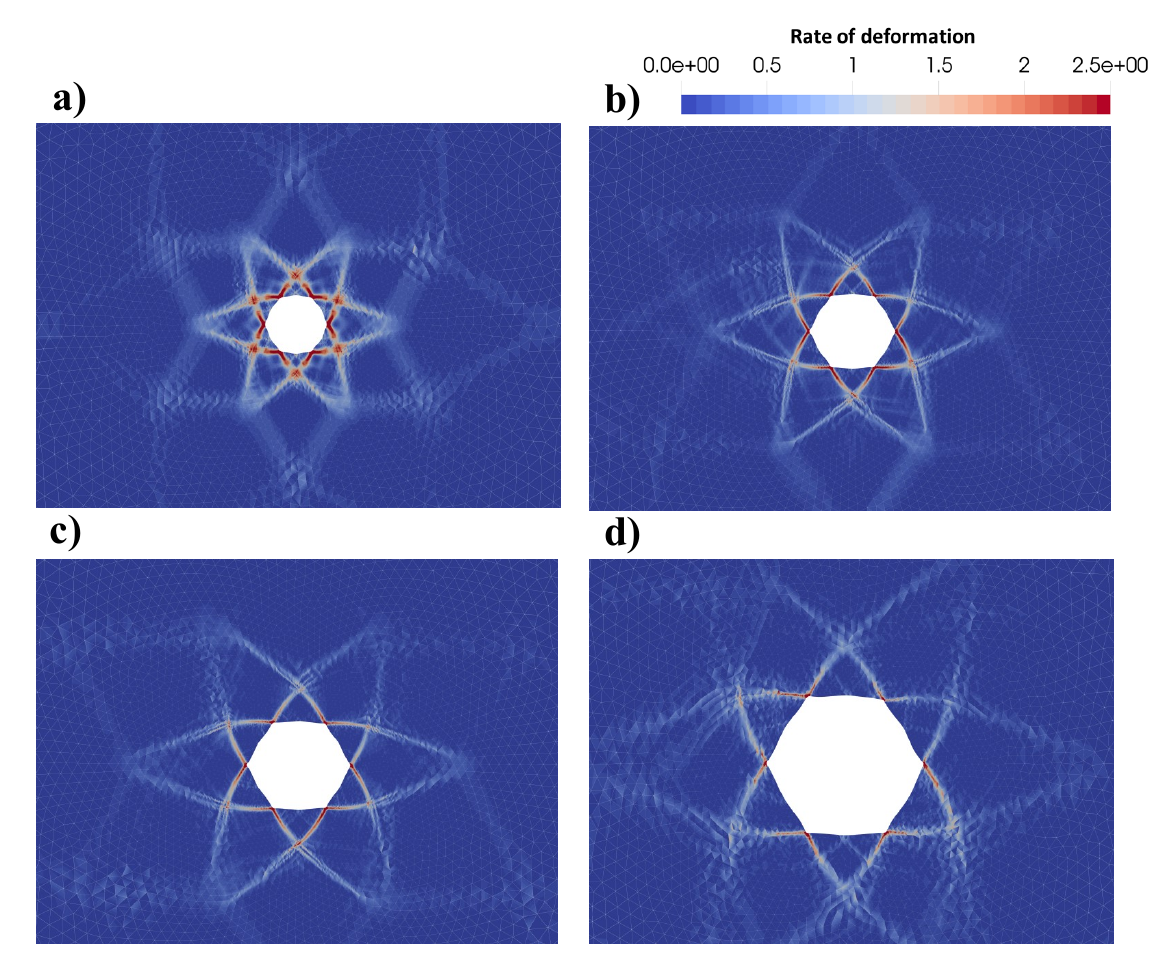}
			\caption{Evolution of equivalent deformation strain rate and void morphology at different stages of the deformation: 
				$\epsilon^{eng}=0.2475\%$,  $\epsilon^{eng}=0.66\%$,    $\epsilon^{eng}=0.99\%$ and $\epsilon^{eng}=1.32\%$.}
			\label{fig:hex_ratedef}
		\end{center}
	\end{figure}

	\begin{figure}[h]
		\includegraphics[scale=0.67]{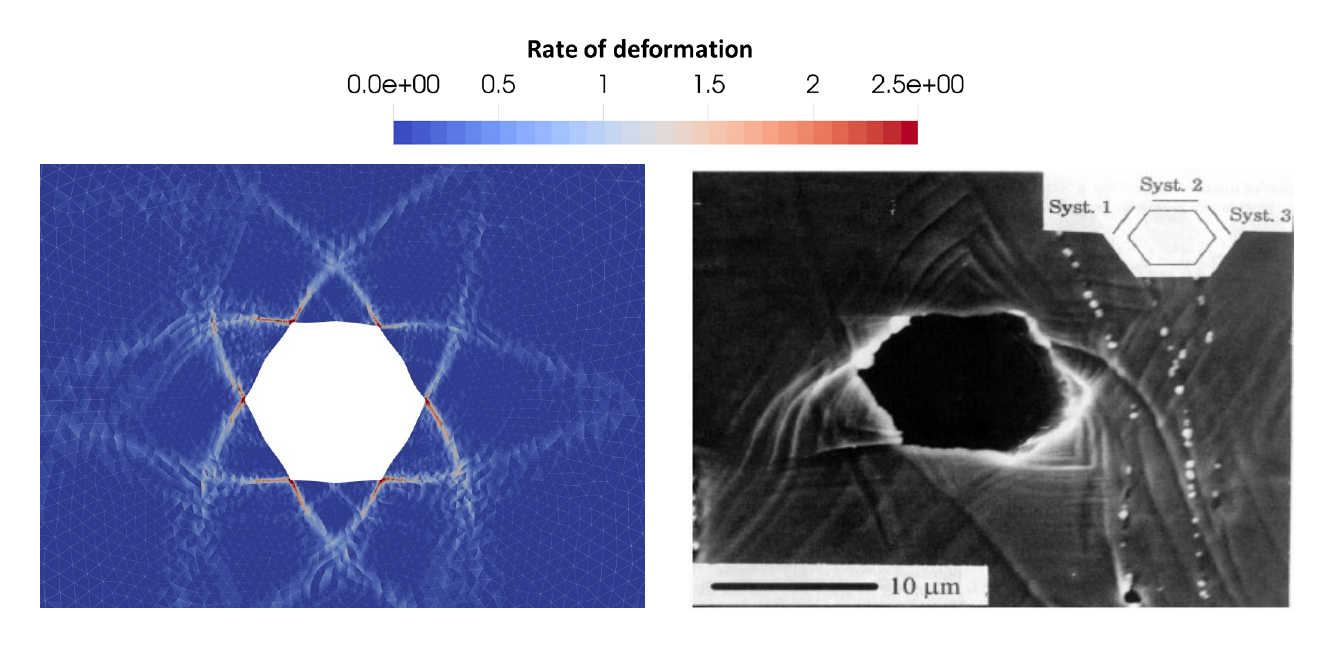}
		\caption{Qualitative comparison between the current model finite element computations and the experimental observations. (Left) Void morphology at $\epsilon^{eng}=0.66\%$. (Right) SEM image of the hexagonal cross-section of the cavity observed on the sample surface, showing the three prismatic slip systems (1, 2, 3) as identified in \cite{crepin1996}.}
		\label{fig:comparaison}
	\end{figure}

	Figure \ref{fig:hex_ratedef} illustrates the evolution of the deformation rate field within the cell and the corresponding changes in void shape at large local deformation for various intervals. As deformation progresses, the void shape  loses its isotropy,   slip bands progressively become dominant in the very surrounding of the void, while the kink bands progressively lose activity and disappear from this vicinity. It is interesting to note that the second stage of the slip line network from the void undergoes opposite evolution and becomes mainly made of kink bands. The persistent activity of the slip bands at elevated local deformation levels drives the transformation of the initially circular void into a hexagonal shape, with its facets aligned along the slip directions. As deformation advances further, as shown from (c) to (d) in Figure \ref{fig:hex_ratedef}, the deformation rate pattern, and therefore the (quasi) hexagonal void undergoes a relatively uniform expansion.  The shear band activity tends to become less intense compared to the period (a) and (b), which can be explained by the increase of the void size, as the strain rate is expected to be roughly proportional to $1/r^2$  \cite{Gurson77,paux2022}.
	
	The final state corresponding to the deformation rate contour derived from Figure \ref{fig:hex_ratedef} (d) is again presented in Figure \ref{fig:comparaison} (a), where we compare the final void shape from finite element simulation results (with its deformation rate) with an experimental observation of a void after plastic deformation reported by \cite{crepin1996}. From a morphological perspective, both images exhibit a similar hexagonal shape of the void. The slip traces that are present in the experimental results suggest that the hexagonal shape results from slip bands around the void similar to those predicted by our numerical simulation.

	Although both the experimental and numerical model consider an HCP crystal with three prismatic systems, it is important to recognize that the loading conditions affecting the void may not be identical. The experimental method employs poly-crystal specimens, making it challenging to determine the exact loading conditions applied in the vicinity of the void and complicates the direct comparison between the simulation and the experiment. Nonetheless, numerical simulations show a strong overall correlation with the experimental results. In particular, it was able to predict the disappearance of the kink bands around the voids under large strain, leading to the same slip activity around the void as predicted by \cite{BORG20076382} without relying on strain gradient plasticity. It is interesting to note that this result cannot be predicted by small strain simulations, where slip and kink bands are strictly equivalent. As the main difference between kink and shear band under large strain is the induced lattice rotation, the evolution of the crystalline orientation in the domain is further investigated.
	
	
	\subsection{Rotation of crystal lattice }
	
	\begin{figure}
		\begin{center}
			\includegraphics[height=13cm]{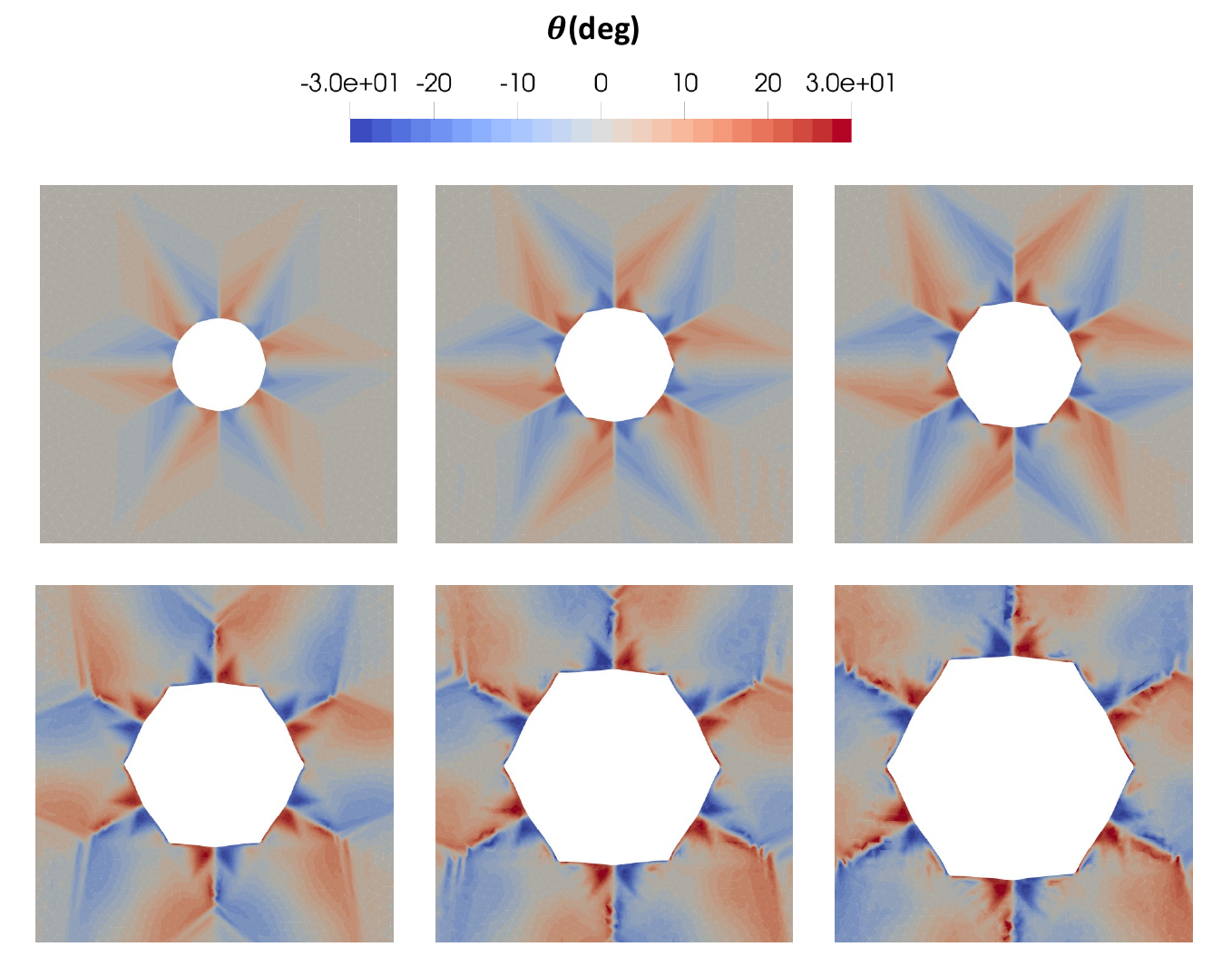}
			
			\caption{Variations in circular cell microstructure orientation $\theta$ (expressed in degrees) during hydrostatic tension loading in the basal plane of an HCP crystal, with the initial crystal orientation set to $\theta = 0^\circ$ at : (a) $\epsilon^{eng}=0.125\%$,  (b) $\epsilon^{eng}=0,2475\%$,  (c) $\epsilon^{eng}=0.33\%$, (d) $\epsilon^{eng}=0.66\%$, (e) $\epsilon^{eng}=0.99\%$ and (f) $\epsilon^{eng}=1.32\%$. The color scale of the contour plot is customized. 
			} 
			\label{fig:hex_orienta}
		\end{center}
	\end{figure} 
	
	As noted earlier, at large deformations, the slip lines around the voids evolve into slip bands, while the kink bands, which are present around the voids at smaller deformations, cease to exist in the very vicinity of the void. One might remark that, at large deformation, the second star-like pattern from the void seems to follow the opposite evolution and is mainly made of kink bands (Figure \ref{fig:hex_ratedef}). Since kink bands play a key role in lattice rotation mechanisms, this transition prompts an investigation into how crystal orientation is affected. The evolution of the lattice rotation around the void, presented in Figure \ref{fig:hex_orienta}, shows that, in regions far from the void, the crystal orientation remains unchanged (close to 0 degrees). 
	We remark that the orientation exhibits a discontinuity at the angles  $\varphi=\pi/6+k\pi/3$ as  the attractor  does. However the discontinuity is more present  near the void, where the deformations are much larger, than near the external boundary. This is in concordance with the  stability analysis of \cite{SSI25}: the 
	distance between the lattice orientation and its attractor decreases significantly for large strains (more than $50\%$)  which are  present near  the void but are  too small  on the external boundary. Moreover, near the void, the orientation distribution formed a star-like pattern with six identical branches, matching the six-star pattern associated with kink bands, which is expected. This pattern is consistent with results at low deformation provided by \cite{gan2006cylindrical}. Within each branch, the orientation varies from $-10^\circ$  to $10^\circ$, with one half of the branch rotating positively and the other half negatively. This bidirectional rotation seems to create a repulsive effect that contributes to the kink bands fading.  Then, the majority of the accumulated plastic strain and slip activity are concentrated in the slip bands. The corner defining the angles of the hexagon forms at the points of contact between the slip bands and the void. In these corner regions, the single crystal experiences a minimal rotation, which allows  the stability of the slip line pattern and therefore the stability of the hexagonal morphology.
	
	As we move closer to the cavity, the highest rotation level of $30^\circ$ degrees appears at the center of each edge of the hexagon, represented by a small half branch  motif mirrored on the opposite side. These points seem to correspond to the location where the kink bands used to touch the void. One half branch shows negative rotation, while the other exhibits positive rotation.
	The disappearance of kink bands at high deformation levels implies that the orientation contour plot shows no significant differences between small and large deformations. The only notable change is the emergence of a second star-like pattern in the final simulation stage, slightly farther from the first, which is consistent with the presence of kink bands in the second stage of the slip line network from the void.
	
	\section{Void growth evolution under uniaxial loading}
	
	\subsection{Setting of the numerical experiment}

\begin{figure} 
		\begin{center}
			\includegraphics[height=9cm]{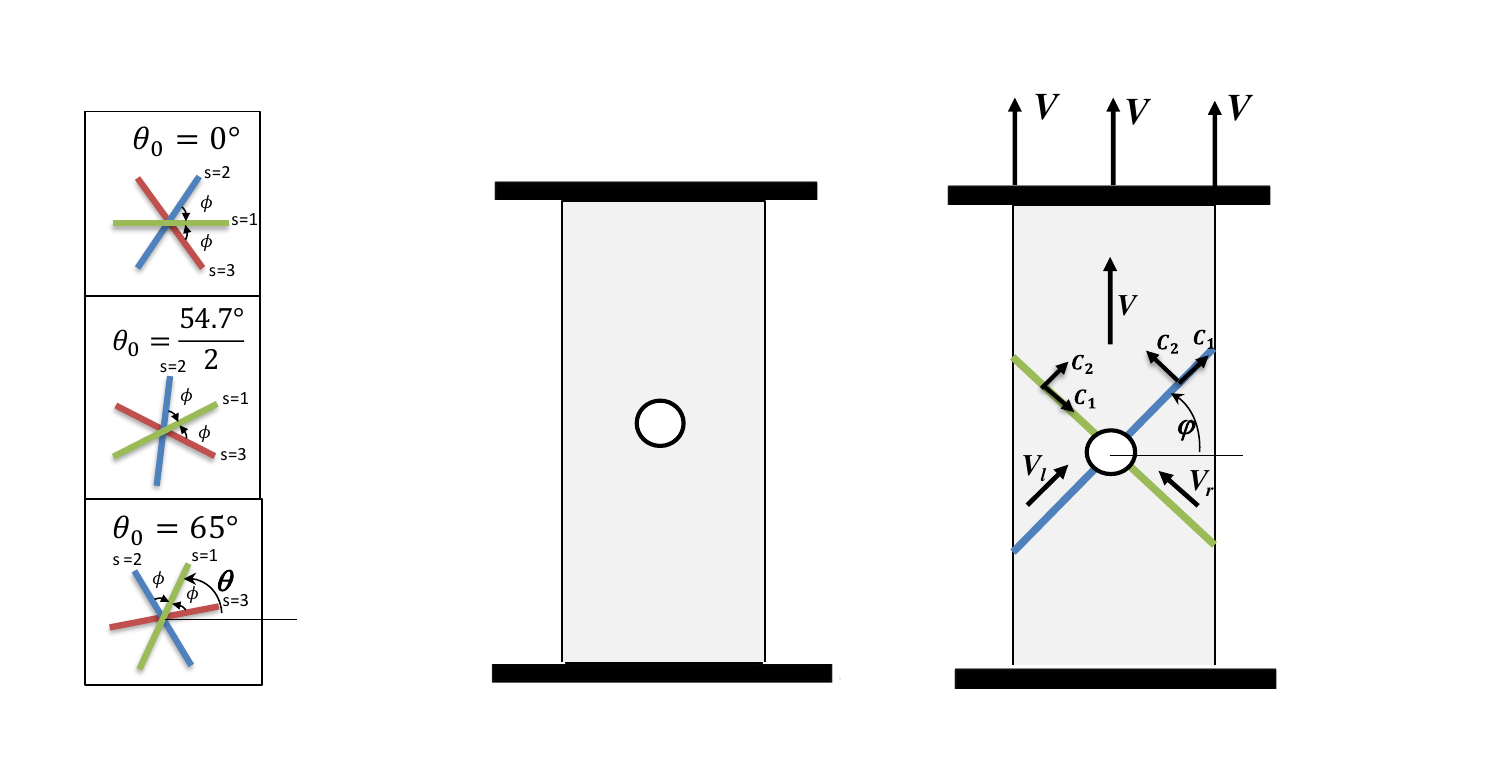}
			\vspace{-1cm}
			\caption{Schematic diagram of a tensile test with four rigid zones. The considered initial crystalline orientations are illustrated on the left. Two  diagonal slip bands forms of angle $\varphi$, indicating the expected localized plastic deformation.} \label{fig:pillar_scheme}
		\end{center}
	\end{figure}

	For FCC crystal under uniaxial loading, the  one-void domain is  represented by a rectangular pillar of dimension $H_0\times L_0$ with $H_0=2L_0$, containing an initially circular void at the center of radius $r_0=\frac{L_0}{30}$ (see Figure \ref{fig:pillar_scheme}). This configuration results in an initial void fraction of $\displaystyle f_{0} = \frac{\pi r^{2}_0}{H_0L_0}= 0.1745 \%$.

	The  pillar's top is animated with a  vertical velocity  $\bV=V\be_2$ (small enough to deal with a quasi-static loading), while the bottom part is kept fixed ($\bV=0$). The  lateral boundary $\Gamma_c(t)$ is stress  free ($\Sb=0$).   The FCC single crystal  has initially the orientation $\theta_0$, i.e. 
	$$\theta(0,x,y)=\theta_0,$$
	with three  choices of $\displaystyle \theta_0 \in \{0^\circ,   \frac{\phi}{2}=\frac{54.7}{2} ^\circ, 65^\circ\}$  to gain an understanding  of the role played by the angle between the loading  axis and the crystal  lattice.

	The engineering  strain is given by 
	$$\epsilon^{eng}(t)=\frac{H(t)-H_0}{H_0},$$
	where $H(t)$  is the current height of the pillar. The material parameters used for this simulation are : density $\rho^{mass}= 16650\,\rm{kg/m^{3}}$, slip resistance for all systems equal to $\tau_c= 220 $ MPa with no hardening,  and a viscosity  $\eta$	chosen, for numerical reasons, to be as small as possible to  ensure that the system is rate-independent (around $1\%$ of $\tau_cH_0/V$).  The  upper velocity $V$ and  the  time interval $[0,T]$ where chosen such that the final engineering  strain is $\epsilon^{eng}_{final}=15\%$.

	\bigskip

	In this test, the top surface of the specimen remains fixed horizontally, ensuring that both the top and bottom surfaces stay parallel and retain the same horizontal coordinates at their extremities, despite significant deformation. The analysis focuses on how shear and kink bands influence the evolution of the void, which is affected by the initial crystallographic orientation of the FCC crystal .
	

	In this study, we analyzed three cases that represent three different initial crystallographic orientations of the  : 
	\begin{itemize}
		\item The first case corresponds to an orientation of $\theta_{0} = 0^\circ$, aligning with one of the three initial slip systems, namely the slip system 1. 
		
		\item The second case involves an orientation of $\theta_{0} = 54.7^\circ/2$, which lies between two slip systems, one at $0^\circ$ and the other at $\theta_{0} = 54.7^\circ$. 
		
		\item The third case considers an  orientation of $\theta_{0} = 65^\circ$, which intends to represent an unparticular orientation.
	\end{itemize}

	\subsection{Evolution of the cavity shape and shear bands localization}
	
\begin{figure} 
		\begin{center}
			\includegraphics[height=20cm]{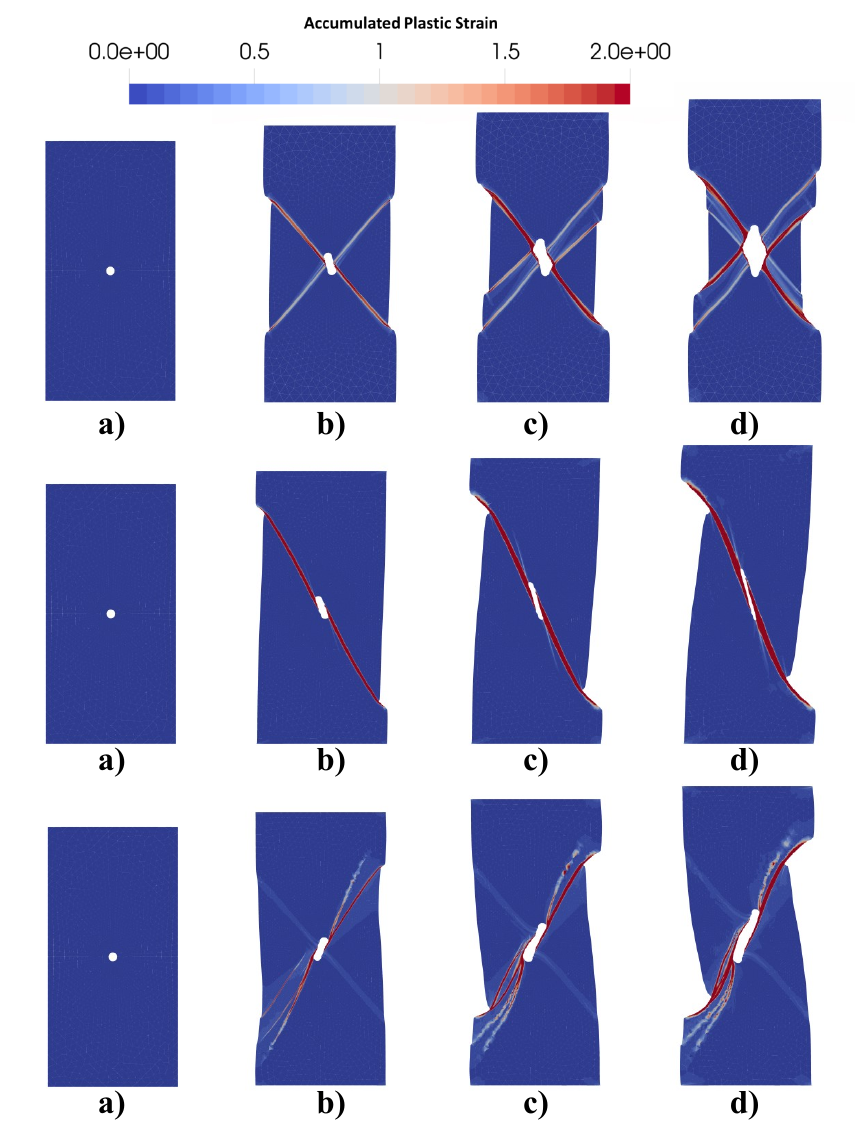}
			\caption{Accumulated plastic strain  $\epsilon^p_a$  ($\dot{\epsilon}^p_a=\vert\bD\vert$) evolution of a FCC pillar under tensile loading with  three initial  crystallographic orientations $\theta_0=0^\circ$ (top), $\displaystyle \theta_0= 54.7^\circ/2$ (middle) and $\theta_0=65^\circ$ (bottom) at   four levels of engineering strains   $\epsilon^{eng}=0,0.05,0.1$ and $0.15$.}
			\label{fig:pillar_plaststrain}
		\end{center}
		
	\end{figure}

\begin{figure} 
		\begin{center}
			\includegraphics[height=20cm]{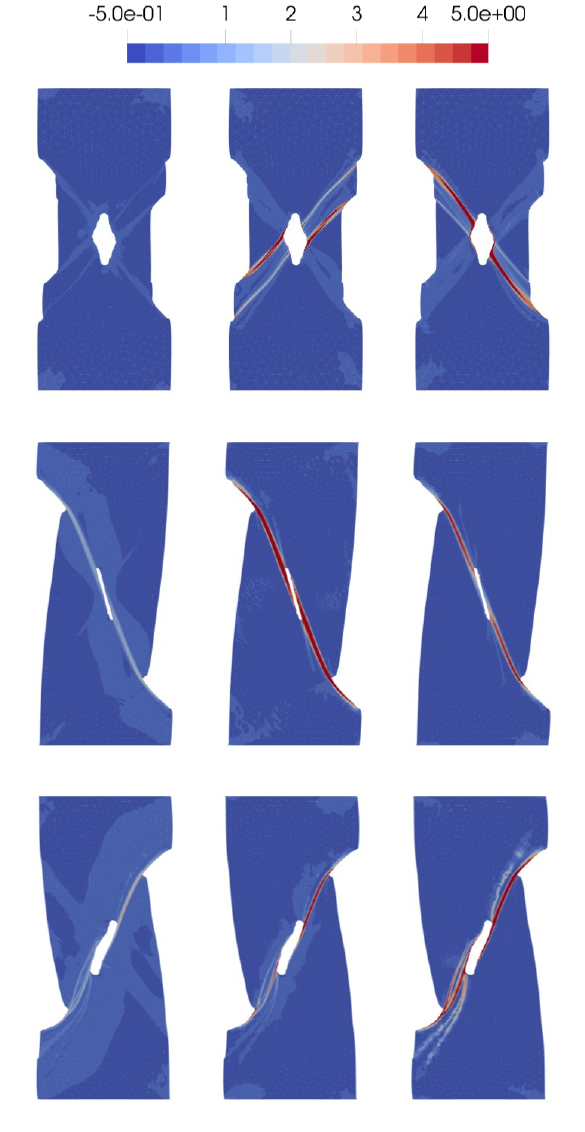}
		\end{center}
		\caption{The spatial distribution of the accumulate slip $\gamma^s_a$  ($\dot{\gamma}^s_a=\vert\dot{\gamma}^s\vert$) on each system at the end ($\epsilon^{engn}=0.15$) of the loading of a FCC  pillar under tensile loading with  three initial  crystallographic orientations $\theta_0=0^\circ$ (top), $\displaystyle \theta_0= 54.7^\circ/2$ (middle) and $\theta_0=65^\circ$ (bottom). The left to right panels respectively give the accumulate slip on system 1, 2 and 3.} \label{fig:slip_act_pillar}
	\end{figure}

\begin{figure} 
		\begin{center}
			\includegraphics[height=20cm]{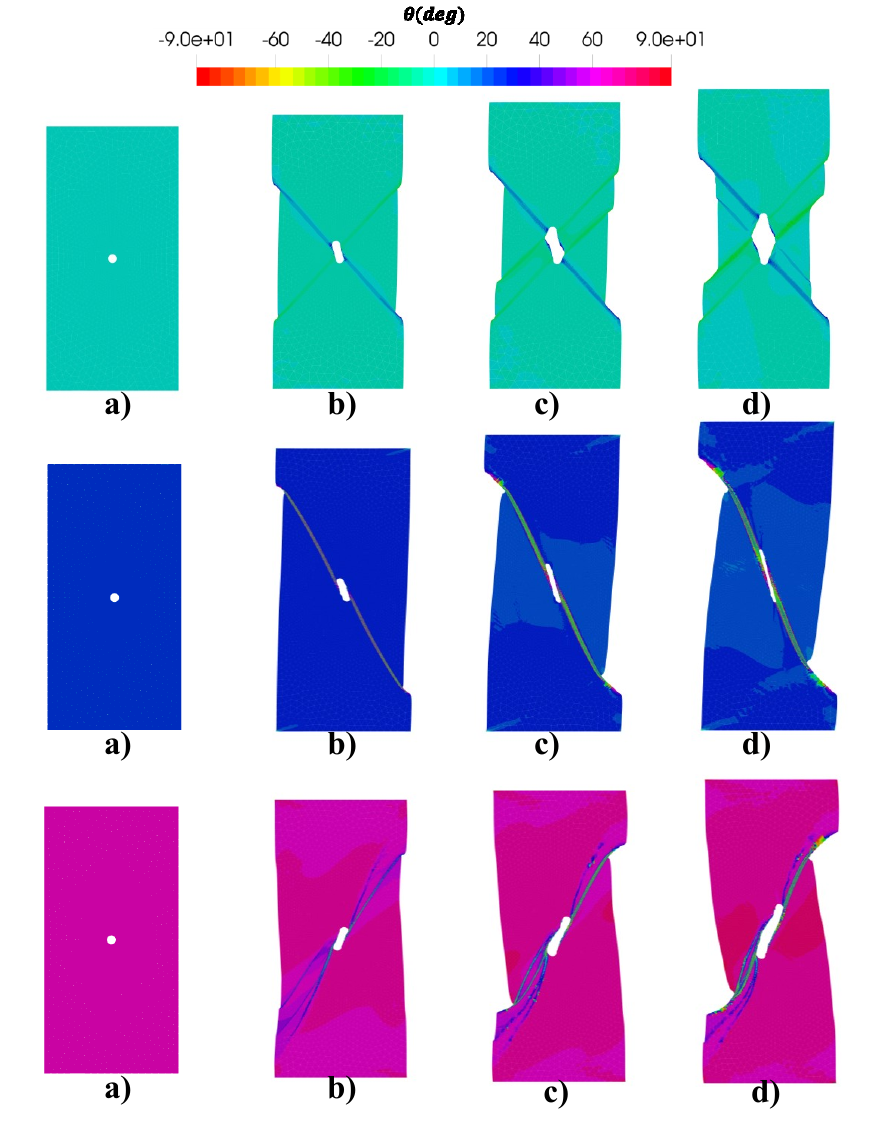}
		\end{center}
		\caption{Lattice orientation evolution of a FCC porous   under tensile loading with  three initial  crystallographic orientations $\theta_0=0^\circ$ (top), $\displaystyle \theta_0= 54.7^\circ/2$ (middle) and $\theta_0=65^\circ$ (bottom) at four levels of engineering strains   $\epsilon^{engn}=0,0.05,0.1$ and $\epsilon^{engn}=0.15$.} \label{fig:pillar_orienta}
		
	\end{figure}

    The Figures \ref{fig:pillar_plaststrain},  \ref{fig:slip_act_pillar} and \ref{fig:pillar_orienta} respectively present the accumulated equivalent strain,  the accumulated slip activity on each system, and the lattice rotation in the three cases at different stages of the tensile test. A first general remark is that the deformation is extremly localized in shear bands, resulting locally in more than $200\%$ accumulated strain, while the global strain is less than $15\%$. This was expected since perfect plasticity (without hardening) is considered. The first case ($\theta_0=0^\circ$, top panels of Figure \ref{fig:pillar_plaststrain}) shows the classical two crossed shear bands at $\varphi=45^\circ$ and $-45^\circ$ (as presented in Figure \ref{fig:pillar_scheme}), while the two other cases ($\theta_0=54.7^\circ/2$ and $65^\circ$, middle and bottom panels of Figure \ref{fig:pillar_plaststrain}) are mostly driven by a unique shear band, showing a high influence of the initial crystalline orientation on the strain field.

	Initial observations show that all voids are distorted under large strain, but that, depending on the initial crystalline orientation, the void either tends to collapse or to grow. This is consistent with the numerical results of \cite{GUO2020102673}, which observed a shape evolution dependency in both crystallographic orientation and stress triaxiality. At high triaxialities, voids tend to grow and coalesce, while at lower triaxialities, distortion, localized shearing, and collapse are more prevalent. Although the effect of stress triaxiality is not analyzed in detail here, it is important to note that the low triaxiality in the uniaxial tensile test favors void distortion, collapse, or closure, while, in the radial growth case discussed earlier, higher triaxiality leads to void growth rather than collapse.

	Continuing from Fig. \ref{fig:pillar_plaststrain}, the scenario of void collapse is seen in the middle panels of the figure, which correspond to an initial angle of $\theta_{0} = 54.7^\circ/2$. Due to the single localized shear band, the circular void progressively flattens and becomes more elongated with continued plastic deformation, eventually forming an almost linear ellipse, with a significantly reduced height (high eccentricity). This scenario is consistent with experimental results obtained with polycrystalline material by Nemcko et al. \cite{nemcko2016effects}, where similar shear band localization is induced by anisotropic voids distribution.
	
	In contrast to the flattening behavior, the void growth, mainly seen in the top panels of the figure ($\theta_0=0^\circ$), occurs through two crossed shear bands, stretching the void along two directions, preventing one dimension from becoming disproportionately small relative to the other.  The double shearing is also observed in the bottom panels ($\theta_{0} = 65^\circ$), but with a dominant shear band, leading to an anisotropic growth of the void, resulting in an elongated void shape, but without collapsing.
	
	These first observations suggest that void shape evolution under uniaxial loading is mainly driven by the presence or absence  of different shear bands around the void. Let's analyse in detail the observed shear bands.

	 First let us remark that in  a slip  band in traction  the velocity gradient  is  $\Lb=2d\bc_1\otimes\bc_2$ where $\bc_1$ and $\bc_2$ are the directions associated to the slip band (see Figure \ref{fig:pillar_scheme}) and $d>0$  for $\varphi>0$ and $d<0$ for $\varphi<0$.  From \eqref{DW} we get that $\bD=d(\bc_1\otimes\bc_2+\bc_2\otimes\bc_1)$ and  $\bW=d(\bc_1\otimes\bc_2-\bc_2\otimes\bc_1)$, which  means that the  rotations are in the clock sense for $\varphi>0$ and in the trigonometric sense for $\varphi<0$.  
	The difference between the  bands  with positive or negative orientation comes from the induced rotations, which therefore are expected to have a huge influence on the evolution of the slip activity. We expect that the lattice in the band will rotate, in the sense given by $\bW$ (i.e. by $\varphi$), till one slip direction is parallel to the slip band, such that, after a while, only one slip system is active. Let us notice that this scenario is valid for one straight shear band with a given constant (in time) orientation $\boldsymbol{c}_1$. This is not always the case, as a complicate process in which the sample shape, the void position, the boundary conditions and the crystalline orientation may interplay to explain  the  existence of  one or several slip bands and their orientations.  
	
	\bigskip
	
	
    For the case $\theta_{0}=0^\circ$  from Figure \ref{fig:pillar_plaststrain}, we deal at the beginning  with  two and at the end with four shear bands. The angles $\varphi$ of all these bands are close to $\pi/4$ or $-\pi/4$. For $\varphi=\pi/4$, the rotation is in the clock sense, hence the first system parallel  with the slip band is $s=2$. This can be seen in  Figure \ref{fig:pillar_orienta} top, with a negative  lattice orientation in the slip band. Moreover, we remark in  Figure \ref{fig:slip_act_pillar} top that the single active slip system is $s=2$. For $\varphi=-\pi/4$, the crystal rotate in the trigonometric sense ($\theta$ positive in the band) and the active slip system is $s=3$ (see Figures  \ref{fig:slip_act_pillar} and \ref{fig:pillar_orienta} top).  At the beginning of the process the presence of the void gives the localization of the two sip bands in the center of the "X", but the void   enlarges a lot during the deformation. This larger void will let another slip band  system appear parallel with one of the incipient slip  bands, but at the end  a fourth  slip band occurs and the shape of the void became symmetric. 	
	
	
	\vspace{10mm}
	
 The case $\theta_{0}= 54.7^\circ/2$ does not follow the schematic representation with four rigid blocks depicted in Figure \ref{fig:pillar_scheme}. The deformation is mostly driven by an unique shear band,  with an orientation varying from  $\varphi=-60^\circ$ to $\varphi=-70^\circ$, approximately. Simultaneously, significant body rotations occur, as can be seen by the rotations of the left and right boundaries of the domain, and confirmed by lattice rotation observation in the middle panels of Figure \ref{fig:pillar_orienta}. The combined influence of shear band deformation and rigid body rotation leads to a complex evolution, hardly explainable by the kind of analysis provided for the $0^\circ$ case, as it cannot be reduced to  a simple  band deformation. Moreover, one can observe in Figure \ref{fig:pillar_plaststrain} that the shear band and the void follow the rigid body rotation, and therefore the slip band does not remain orientated at $\varphi=-60^\circ$ during the evolution of the considered geometry. This competition between shear band deformation and rigid body motion prevents from the stabilization of the lattice orientation on a slip band, leading to multi-slip deformation during all the tensile test, with mostly slip activity from the two  systems 2 and 3 (Figure \ref{fig:slip_act_pillar}, middle panels). One might remark that the rigid body rotation might be a consequence of the imposed null lateral displacement at the top and at the bottom of the pillar, making a unique shear band unable to perform the necessary deformation alone, as it induces non null lateral displacement, as schematized  in Figure \ref{fig:pillar_scheme} (velocity vectors $\boldsymbol{V}_l$ or $\boldsymbol{V}_r$). Different boundary conditions may lead to different shear band localization.
	\vspace{10mm}

	
	Finally, a configuration closed to the $54.7^\circ/2$ case is obtained for $\theta_{0}= 65^\circ$, with a dominant shear band initially oriented at approximately $\varphi=50^\circ$ and significant rigid body rotation. Again, this leads to multi-slip activity, with dominant activity of the  systems 2 and 3 (bottom panels of Figure \ref{fig:slip_act_pillar}), and no stabilization of the lattice rotation (bottom panels of Figure \ref{fig:pillar_orienta}). Contrary to the case $\theta_{0}= 54.7^\circ/2$, a second shear band with lower intensity  appears at approximately $\varphi=-50^\circ$. These combined shear band deformations lead to a scattering of the dominant shear band, whose past plastic deformation is transported by the second shear band deformation, leading to the presence of layers of past shear band deformations (Figure \ref{fig:pillar_plaststrain}, bottom panel d). This scattering looks like the trace of plastic slips experimentally observed by Crépin et al. \cite{crepin1996} (see Figure \ref{fig:comparaison}), and suggests sequential activation of the two shear bands.

	
	
	\section{Conclusions}
	
	In this  paper, the large deformation  growth  of voids in single crystals (HCP and FCC)  was investigated.  To this end, an Eulerian crystal plasticity 2D model with three slip systems was employed and numerically solved in finite element simulations.  The  voids' deformation  was analyzed under two types of loading: (i) radial and (ii) uni-axial.   In both cases, the voids, which  have an initial circular shape, undergoes significant shape distortion, and shear bands appear, where the accumulated plastic strain is very large  ($>200\%$), even for moderate global engineering strains ($<15\%$).
	
	
	In  case (i), the onset of deformation shows the formation of a slip and kink bands network (already observed in \cite{paux2022}). This network evolves under large strain, with vanishing of the kink bands around the voids, leading only the slip bands drive the evolution of the shape of the void. It ultimately leads to an   {\em hexagonal shape} of the void, consistently with experimental observations  (see for instance \cite{crepin1996}). In the meantime, significant lattice rotations are observed around the void, in accordance with the kink band pattern, explaining the loss of symmetry between kink and slip bands in this vicinity. The final orientation of the crystal  around the  void is given with 6 lines of discontinuity, with rotation up to $30^\circ$, while, far from the void, accumulated deformations are not large enough to induce significant lattice rotations.

	In case (ii) we have considered three initial orientations of the crystal with respect to the  loading axis.  We get two different scenarios   of the void evolution. Only one scenario corresponds to the  classical one  involving four blocks separated by slip bands with only one active slip system in each band. In this scenario  the initial circular void has a final rhomboidal  shape. The other one exhibits only one shear band with a variable (in time)  orientation, and   with two active slip systems.   The final shape of the void is a thin rectangle oriented almost parallel with the shear band. In this  scenario we expect that the void will collapse in a crack and an auto-contact will occur.
	
	These results demonstrate the ability of the Eulerian approach to model crystal plasticity under very large strain, with shear band localization and significant crystalline rotations. Moreover, the obtained results provide a better understanding of the micro mechanisms leading to the evolution of the porosity under large strain, and the importance of the complex crystal plasticity related phenomenon on the evolution of the shape of the voids at the crystalline scale, which are of significant importance for the onset of ductile fracture.
	
	
	
	\bibliography{Void_submission_Arxiv}
	
	\begin{appendices}
		\part*{Appendix}
		
		\section{Numerical scheme used for the simulations} \label{Ann:Numerical}
		
		The main goal of this section is to recall the 
		Eulerian numerical strategy proposed in \cite{Cazacu2010-hi} for  the  simplified 2D  rigid-visco-plastic  crystal   model used in this  paper.  
		
		The use of  a  time implicit (backward) Euler scheme   for time discretization 
		gives  a set of nonlinear equations for the velocities $\bu$ and lattice orientation  $\theta$.  
		At each  iteration in time,     an iterative  algorithm is developed to solve these nonlinear equations. Specifically, a mixed finite-element (FE) and  Galerkin discontinuous (GD) strategy is proposed. The variational  formulation  for the velocity field is discretized using the finite element method, while a Galerkin discontinuous method with an upwind choice of the flux is adopted for solving the hyperbolic equations that describe the evolution of the lattice orientation. It is to be noted that in the case of the rigid-viscoplastic model  studied in this work, additional difficulties arise from the non-differentiability of the plastic terms.  That means that we cannot simply make use of the finite element techniques developed for Navier-Stokes fluids (see for instance \cite{P,tem}).   To overcome these difficulties the iterative  decomposition-coordination formulation coupled with the augmented Lagrangian method (introduced in   \cite{GlLT,FoGl})  was  modified. 
		The reason for this modification  is that, in the crystal model there is  non co-axiality between the stress deviator and the rate of deformation in contrast with the von-Mises model for which the original method was proposed (see  \cite{GlLT,FoGl}).  
		This type of algorithm permits also to solve alternatively,  at each iteration,  the equations  for the velocity field and   for the unit vectors that define the lattice orientation.  
		
		For vanishing   viscosity,  the adopted  visco-plastic model  contains as a limit case the inviscid Schmid law.  Even that Schmid model  is very stiff,  for small viscosities (as for metals)  and moderate strain rates the iterative decomposition coordination formulation coupled with the augmented Lagrangian method  works very well and no instabilities are presents.  However,  for  small viscosities  and large velocities (more then $100$m/s)   with large Reynolds numbers (Re $ > 1000$)  the above algorithm is converging very slowly. In this last case,  other techniques, coming from  gas dynamics, have to be included.

		\bigskip
		If the Eulerian  domain $\DD$  has time variations  then the above algorithm has been  adapted  to an ALE (Arbitrary Eulerian-Lagrangian) description of the crystal evolution.  In this case is more convenient to have the same finite FE and GD meshes. This avoid the interpolation of the lattice orientation  on the deformed mesh.   As a matter of fact, the numerical algorithm proposed here deals only with a Stokes-type problem at each time step and the  implementation of the Navier-Stokes equations in an ALE formulation is rather standard  (see for instance \cite{hughes1981lagrangian,maury1996characteristics,maronnier2003numerical,duarte2004arbitrary}).
		
		For the sake of simplicity we shall present  here the numerical scheme for the simplified  $N=3$ slips systems in 2D, described  in section 2. 
		
		\subsection{Time discretization}
		
		Having in mind the  initial and boundary value problem, stated in   Section {BVP},  let $\Delta t$ be the time step and let us denote by $\bu^k$,
		$\bs^k$, $p^k$ and  $\theta^k$ the values of the unknowns $\bu(k\Delta
		t)$, $\bs(k\Delta t)$,  $p(k\Delta t)$, and  $\theta(k\Delta t)$. Suppose that we have computed all these variables at time
		$t=(k-1)\Delta t$. Let also denote  by ${\bf V}^{k}={\bf V}(k\Delta t),  {\bf S}^{k}= {\bf S}(k\Delta t),  \theta_{in}^k=\theta_{in}(k\Delta t)$ the  boundary conditions at $t=k\Delta t$.

		The time implicit (backward) Euler scheme   for the field equations of the  initial and boundary value problem 
		gives the following nonlinear equations for
		$\bu^{k}$, $\bs^{k}$, $\theta_s^{k}$ and $\rho^{sk}$
		\begin{equation} \label{mbl-k}
			\rho^{mass}\frac{ \bu^{k} -\bu^{k-1}}{ \Delta t}+ \bu^{k} \cdot \nnabla \bu^{k}
			-  \bfdiv \bs'^{k} + \nabla p^k= \rho \ff  \quad \hbox{ in }
			\DD,
		\end{equation}
		\begin{equation}\label{inc-k}
			{\rm div} (\bu^{k}) =0  \quad
			\hbox{ in } \DD,
		\end{equation}
		\begin{equation}\label{bT-k}
			\frac{  \theta^{k}- \theta^{k-1}}{\Delta t}+   \bu^{k}\cdot \nnabla \theta^{k} =\frac{1}{2}\left(\sum_{s=1}^{N} \dot{\gamma}_s^k - (\frac{ \partial v^k_1}{ \partial x_2}-\frac{ \partial v^k_2}{ \partial x_1})\right),
		\end{equation}
		\begin{equation}\label{DS-k}
			\D(\bu^{k})= \sum_{s=1}^{N}\dot{\gamma}_s^k\M_s^{k}, \quad \quad  \dot{\gamma}_s^k=\dfrac{1}{\eta_s}\left[1-\dfrac{\tau^{sk}_0}{\left\vert\bs^{k}:\M_s^{k}\right\vert}\right]_+\left(\bs^{k}:\M_s^{k}\right),
		\end{equation}
		while the boundary conditions read 
		\begin{equation} \label{bc-k}
			\bu^k = {\bf V}^k \; \hbox{on} \; \Gamma_v, \quad \bs^k \bn = {\bf
				S}^k
			\; \hbox{on} \;
			\Gamma_s,  \quad  \theta_s^k =\theta_{in}^k,       \quad   \hbox{on} \;  
			\partial_{in} \DD.
		\end{equation}

		\subsection{The algorithm at each time step}
		
		Let us fix the iteration in time, $k$.  An iterative  decomposition-coordination formulation coupled with the augmented  method (see \cite{GlLT,FoGl}) will be  adapted here for the  crystal plasticity model.  This type of algorithm permits  to solve alternatively,  at each iteration $n$,  equations
		(\ref{mbl-k}-\ref{DS-k})  for the velocity field and  (\ref{bT-k}) for the lattice orientation.
		The convergence is achieved  when  the difference between
		$\bu^{k,n}, \bs^{k,n}, \theta^{k,n}_s$ and $\bu^{k,n-1}, \bs^{k,n-1}, \theta^{k,n-1}_s$ is small enough.
		
		In order to describe the algorithm let $r>0$ be  the augmented Lagrangian step and let  $\DD$ be discretized by using a family of triangulations $({\cal
			T}_h)_h$ made of finite elements (here $h>0$ is the discretization parameter representing
		the greatest diameter of a triangle in  ${\cal T}_h$).   We  denote  by $V_h$ the FE space for the velocity field $\bu^{k}$, by $W_h$ the FE space for the pressures field $p$ and by $Q_h$ the Galerkin discontinuous   space for the stresses deviators $\bs'$,  for lattice orientations $\theta$ and dislocation densities $\rho_s$).

		We put   $\bu^{k,0}=\bu^{k-1}, \bs'^{k,0}=:\bs'^{k-1}, \theta^{k,0}_s=\theta^{k-1}_s,  \dot{\delta}_s^{k,0}= \dot{\gamma}_s^{k-1}$ and we suppose that   $\bu^{k,n-1}, \bs'^{k,n-1}$  and $\theta^{k,n-1}_s,\dot{\delta}_s^{k,n-1}$ are known.  

		\bigskip
		
		{\bf Step 1.)} The first step consists in solving the following linear equation of Stokes type for the velocity field $\bu^{k,n}$  and the pressure $p^{k,n}$:
		\begin{eqnarray} \label{mbl-kn}
			\rho^{mass}\left(\frac{ \bu^{k,n} -\bu^{k-1}}{ \Delta t}+ \bu^{k,n-1} \cdot \nabla \bu^{k,n-1}\right)
			-  \bfdiv \left(r\D(\bu^{k,n})\right) + \nabla p^{k,n}=  \nonumber \\    \bfdiv  \tilde{\bs}^{k,n-1} + \rho^{mass} \ff,
			\quad  \quad  \quad 	{\rm div} (\bu^{k,n}) =0, \end{eqnarray}
		with the boundary conditions
		$$	\bu^{kn} = {\bf V}^k \; \hbox{on} \; \Gamma_v, \quad  \left(r\Db(\bu^{k,n}) - p^{k,n}\I +  \tilde{\bs}^{k,n-1} \right) \bn={\bf	S}^k \quad \hbox{on} \quad \Gamma_s.$$
		where we have denoted by 
		$$ 
		\tilde{\bs}^{k,n-1}=\bs'^{k,n-1}- r\sum_{s=1}^{N}  \dot{\delta}_s^{k,n-1} \M_s^{k,n-1}.
		$$
		The above problem  is a standard one  in fluid mechanics and there exists many techniques to solve it.  In all the computations presented in this work, we have used a Lagrangian formulation   with  a  [continuous P2, continuous P1]   choice for  finite element spaces  $(V_h, W_h)$  associated to the  velocities  and pressures.
		
		\bigskip
		
		
		{\bf Step 2.)} The second step consists in finding  the decomposition of  the global rate of deformation $\D(\bu^{k,n})$,  into the slip rates  $\dot{\gamma}_{s}^{k,n} \in Q_h$, according to (\ref{DS-k}). 
		This can be done using the  analytic formula, which provide directly the slip rates $\dot{\gamma}_{s}^{k,n}$ from the expression of  $\D(\bu^{k,n})$ and of the yield limit $\tau^{sk,n-1}_0$, see \cite{smiri:tel-05010824}.

		\bigskip
		
		Note that   the finite element spaces for the velocity fields   $\bu^{k,n}$ and that for the  slip rates $\dot{\gamma}_{s}^{k,n}$ cannot be chosen independently.   For instance if $V_h$ = [continuous P2] then we have to choose  $Q_h$=[discontinuous P1].  This is the choice in all simulations presented in this work. 
		
		\bigskip
		
		{\bf Step 3.)}  We introduce   now  the slip rate multipliers $\dot{\delta}_s^{k,n} : \DD \to \R$, belonging to  $Q_h$,     computed   according to the decomposition-coordination formulation
		coupled with the augmented  method for each slip system:
		\begin{equation}\label{flow_rul-kn}
			\dot{\delta}_s^{k,n} = \dfrac{1}{\eta_s +r}\left[1-\dfrac{\tau^{sk,n-1}_0}{\vert\bs^{k,n-1}:\M_s^{k,n-1} +r \dot{\gamma}_{s}^{k,n}\vert} \right]_+ (\bs^{k,n-1}:\M_s^{k,n-1}+r \dot{\gamma}_{s}^{k,n}). 
		\end{equation}
		Then the  deviator stress field is updated:
		\begin{equation}\label{stress-kn}
			\bs'^{k,n} =  \bs'^{k,n-1} +  r\left( \D(\bu^{k,n}) - \sum_{s=1}^{N} \dot{\delta}_s^{k,n}  \M_s^{k,n-1} \right).
		\end{equation}
		
		\bigskip

		{\bf Step 4.)} In this  step, we compute the  lattice orientation  from the linear hyperbolic equation for $\theta^{k,n}$
		\begin{equation}\label{theta-kn}
			\frac{  \theta^{k,n}- \theta^{k-1}}{\Delta t}+   \bu^{k,n}\cdot \nnabla \theta^{k,n} =\frac{1}{2}\left(\sum_{s=1}^{N} \dot{\gamma}_s^{k,n} - (\frac{ \partial v^{k,,n}_1}{ \partial x_2}-\frac{ \partial v^{kn}_2}{ \partial x_1})\right),
		\end{equation}
		with the boundary conditions 
		$$  \theta^{k,n} =\theta_{in}^k,      \quad   \hbox{on} \;  \partial_{in} \DD. $$

		To solve the linear system (\ref{theta-kn})   we have adopted here a GD  strategy with an "upwind" choice of the flux.  In the numerical applications presented in this paper we have chosen the finite volume mesh to be  the finite element triangulation ${\cal T}_h$, and the  finite volume  space to be  $Q_h$=[discontinuous P1].

		\subsection{The algorithm in an ALE  method} 
		
		If the   domain $\DD$  occupied by the single crystal (or poly-crystals) varies in  time,  then an arbitrary Eulerian-Lagrangian  (ALE )  description was adopted.  We want to point out here  how the above algorithm have to be changed if  it is used  coupled with an   ALE method.  
		For that we have to have in mind that  the passage from time iteration $k-1$ to $k$ involves the   frame velocity $\bu_{fr}^{k-1}$.  Since  in the ALE formulation $\bu_{fr}^{k-1}\cdot \bn=\bu^{k-1}\cdot \bn$  the income boundary $\partial_{in} \DD(t)$ is always empty. There are   only three equations from the above algorithm which  have to be changed:  equation (\ref{mbl-kn}) with 
		\begin{eqnarray} \label{mbl-knALE}
			\rho^{mass}\left(\frac{ \bu^{k,n} -\bu^{k-1}}{ \Delta t}+ (\bu^{k,n-1}-\bu_{fr}^{k-1}) \cdot \nabla \bu^{k,n-1}\right)
			-  \bfdiv \left(r\D(\bu^{k,n})\right) + \nabla p^{k,n}=  \nonumber \\    \bfdiv  \tilde{\bs}^{k,n-1} + \rho^{mass} \ff,
			\quad  	{\rm div} (\bu^{k,n}) =0, \end{eqnarray}
		equation (\ref{theta-kn}) with 
		\begin{equation}\label{theta-knALE}
			\frac{  \theta^{k,n}- \theta^{k-1}}{\Delta t}+   (\bu^{k,n}-\bu_{fr}^{k-1})\cdot \nnabla \theta^{k,n} =\frac{1}{2}\left(\sum_{s=1}^{N} \dot{\gamma}_s^{kn} - (\frac{ \partial v^{k,n}_1}{ \partial x_2}-\frac{ \partial v^{k,n}_2}{ \partial x_1})\right).
		\end{equation}

		\section{Remeshing procedure} 
		The quality of the simulation results in mesh based approaches very much depends upon the characteristics of the mesh. A poor mesh quality impacts the computational efficiency, increases the computational time and may lead to unstable solutions\cite{MR3043640}. Meshing can be either uniform or non-uniform.
		Adaptive meshing is a type of non-uniform mesh scheme that is widely  used in FEM based approaches see \cite{sedighiani2021large}. It is characterized by a mesh density that varies across different regions. 
		We employ a remeshing technique to address mesh distortion issues in crystal plasticity simulations. This method involves replacing distorted meshes with new, undistorted ones  \cite{sartorti2024remeshing}. The variables from the deformed configuration are transferred to the new mesh using a nearest-neighbor mapping algorithm. The simulation is then restarted with the initial state set based on the most recent deformation state. During large deformations, the aspect ratio of the elements, defined as the ratio between element size in the stretching versus compression directions, can become excessively large. This can introduce errors and hinder strain localization. To mitigate this, a multi-step mesh refinement strategy is applied at each remeshing stage, adjusting the mesh density as the deformation progresses. The key idea is to maintain a constant number of elements in the compression direction while increasing the number of elements in the stretching direction to preserve a near-cubic element shape. This gradual mesh refinement enhances simulation resolution during deformation while minimizing information loss and avoiding a significant increase in computational cost (see \cite{resk2009adaptive}).

		\begin{figure}
			\center
			\makebox[\textwidth][c]{\includegraphics[width=1\textwidth]{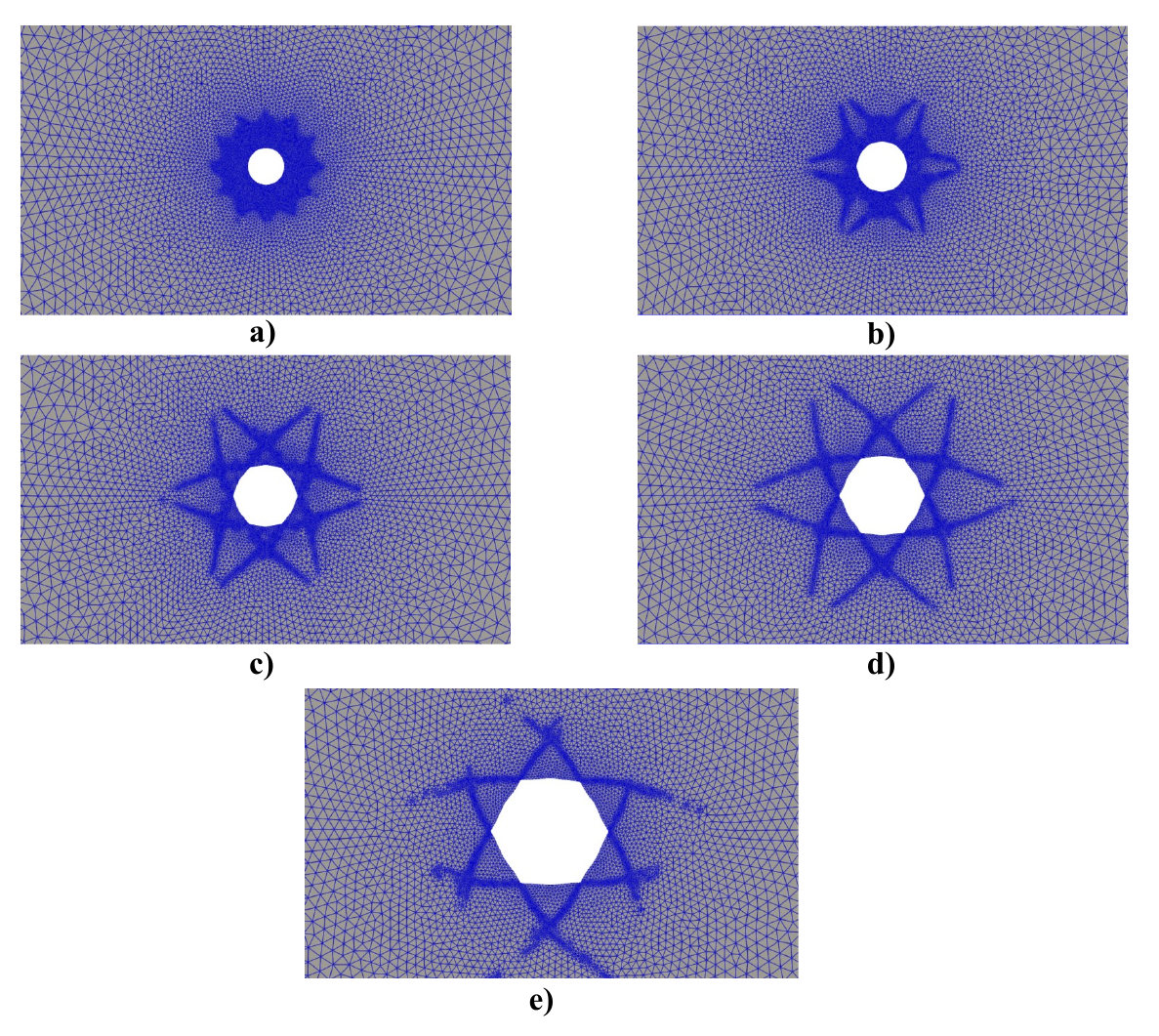}} 
			\caption{Adaptive meshing evolution for a disc unit cell with a single circular void at its center (monocrystal HPC) under radial loading simulations (section 3). The mesh distributions correspond to  (a) $t=T/80$ ($\epsilon^{eng}=0.0165\%$), (b) $t=T/10$ ($\epsilon^{eng}=0.132\%$), (c) $t=T/5$ ($\epsilon^{eng}=0.264\%$), $t=2T/5$ ($\epsilon^{eng}=0.528\%$) and (d) $t=4T/5$  ($\epsilon^{eng}=1.056\%$). }  \label{remeshvoid} 
		\end{figure}

		In our adaptive meshing approach for single crystal and polycrystal simulations, we apply three combined criteria: strain gradient, orientation gradient, and accumulative plastic strain. 
		\begin{itemize}
			\item Velocity Gradient Criterion : The strain gradient serves as a crucial indicator of regions experiencing significant deformation. Areas with steep strain gradients often correspond to locations of localized plasticity, such as near grain boundaries or in regions undergoing shear band formation. By refining the mesh in these high-gradient areas, we ensure that the simulation can accurately model the material's response to applied loads and capture critical features of the deformation process.
			
			\item Orientation Gradient Criterion: The orientation gradient is equally important in polycrystal simulations, as it reflects the variation in crystallographic orientation across the material. Changes in orientation can significantly influence mechanical behavior, particularly in materials exhibiting anisotropic properties. By monitoring the orientation gradient, we can identify regions where the crystal orientations change rapidly, necessitating a finer mesh to resolve the interactions between grains and accurately capture the evolution of microstructural features.
			
			\item Accumulative Plastic Strain Criterion: This additional criterion targets areas of high plastic strain to minimize mesh distortion during ongoing deformation. Finer meshing in these zones helps maintain element integrity, reduce numerical artifacts, and enhance stability.
		\end{itemize}
		Combining these criteria allows for a more nuanced adaptive meshing strategy. In regions where both the strain and orientation gradients are high, the mesh density is significantly increased, ensuring precise resolution of complex interactions and behaviors. Conversely, in areas where both gradients are low, the mesh can be coarsened to optimize computational efficiency without sacrificing accuracy.
		
		These criteria adaptive meshing approach not only enhances the fidelity of our simulations but also minimizes computational costs by focusing resources where they are most needed. As the simulation progresses, the mesh dynamically adapts to the evolving deformation field and microstructural characteristics, resulting in improved simulation accuracy for crystalline materials under large deformations.

		While this multi-criteria strategy generally captures complex deformation behaviors in polycrystalline materials and improves accuracy, using all three criteria simultaneously is not always necessary or efficient. For instance, when orientation gradients are weak, a dual-criterion approach using strain gradient and accumulative plastic strain may suffice. Additionally, when strain gradient and accumulative plastic strain are both significant within the same region, the strain gradient criterion alone may be used to capture the essential deformation characteristics.

		To illustrate the effectiveness of our adaptive meshing strategy, mesh refinement patterns derived from various simulations discussed in the preceding  sections are presented. The figures illustrate the adaptive meshing process across different simulations of single-crystal structures under various loading conditions and strain levels. Each figure clearly demonstrates how the mesh adjusts to capture deformation characteristics specific to the crystal type, initial orientation, and applied strain.

		In Figure \ref{remeshvoid} which coresponds to the section 3, the mesh begins with a dense refinement around the void, forming two overlapping star-like structures: one defined by kink bands and the other by slip bands. As deformation progresses, the kink-band structure gradually diminishes, leaving only the star pattern formed by shear bands. This evolution in mesh refinement effectively captures the localized deformation and the transition in dominant band structures surrounding the void.
		
\begin{figure}
			\center
			\includegraphics[width=0.6\textwidth]{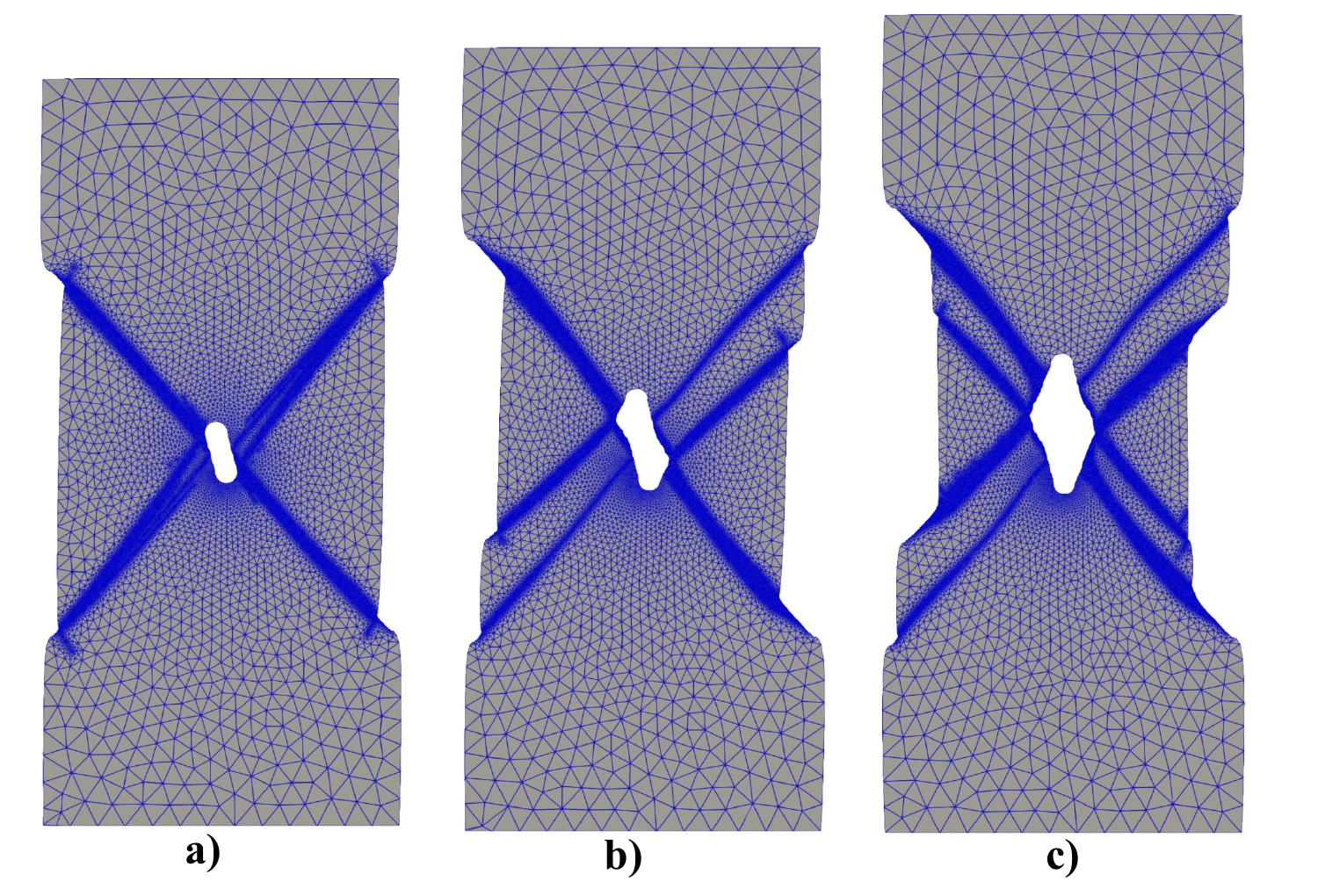} 
			\includegraphics[width=0.6\textwidth]{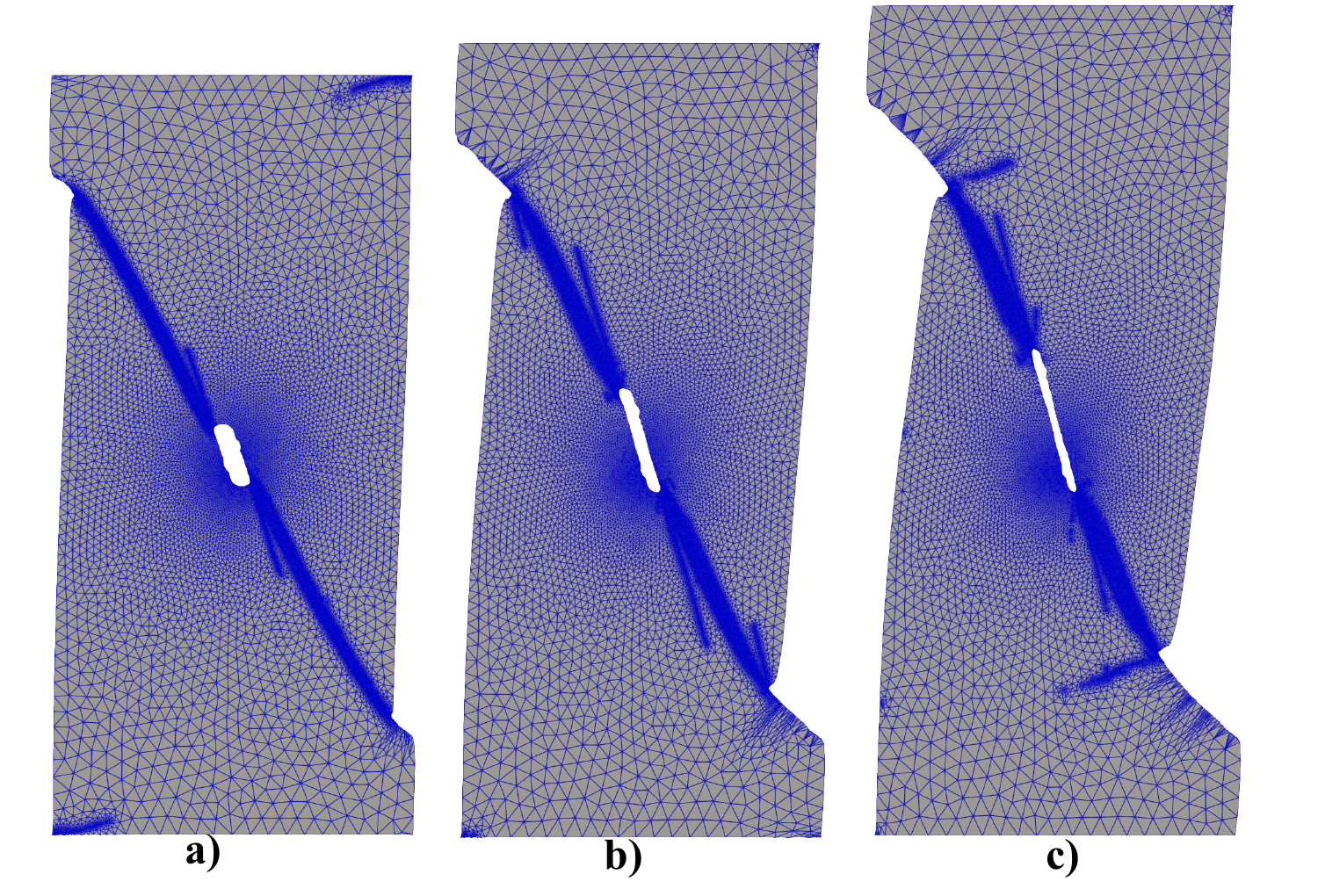} 
			\includegraphics[width=0.6\textwidth]{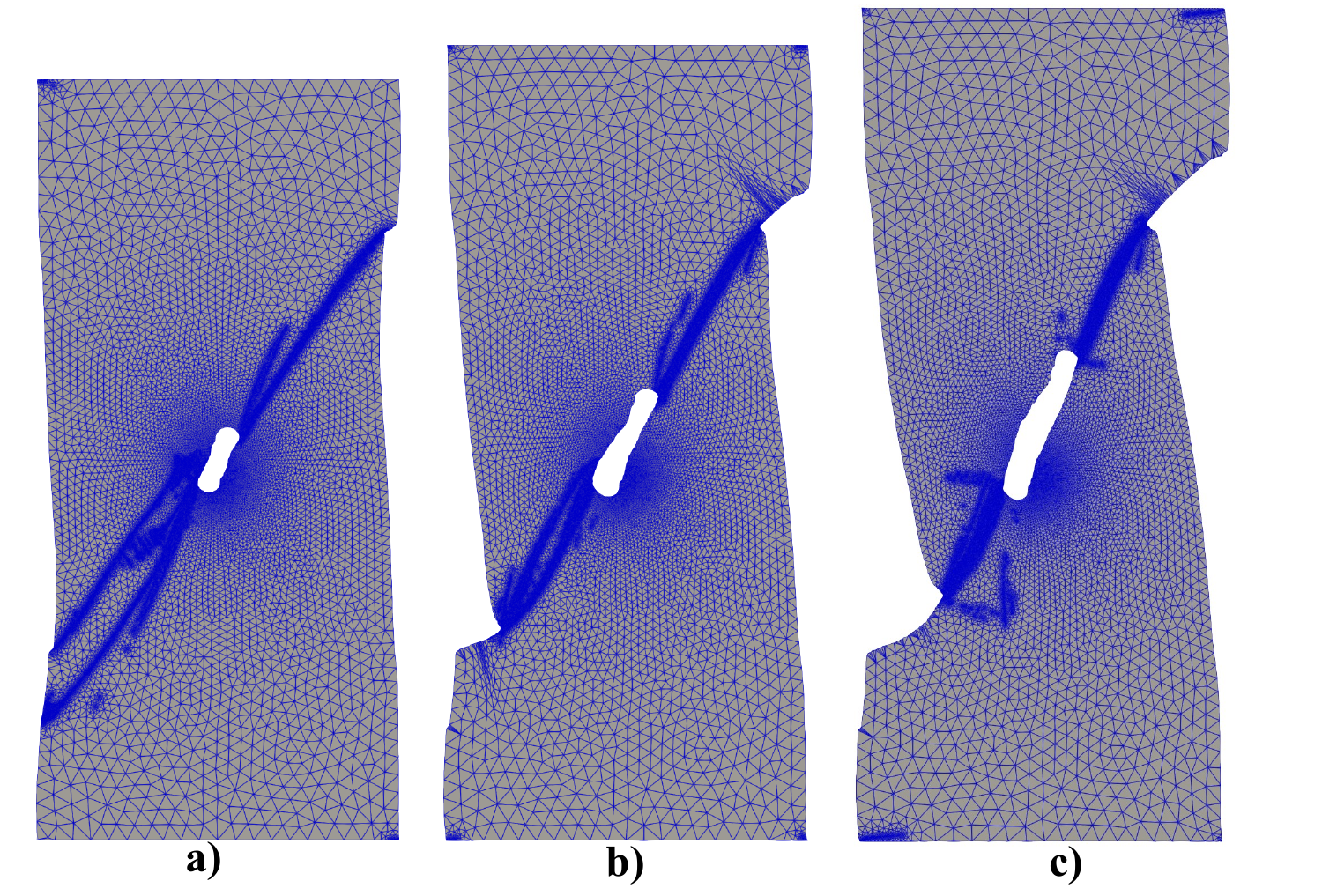} 
			\caption{Adaptive meshing evolution of a FCC porous   under tensile loading (see section 4),  with initial crystallographic orientations $\theta_0=0^\circ$ (top),  $\theta_0=\frac{54.7}{2}^\circ$ (middle) and $\theta_0=65^\circ$ (bottom) at three levels of engineering strains : (a) $\epsilon^{eng}=0.05$, (b) $\epsilon^{eng}=0.1$ and (c) $\epsilon^{eng}=0.15$. }  \label{remeshFCC0}
		\end{figure}

		Figures \ref{remeshFCC0} illustrates meshing in an FCC porous  under tensile loading with different crystallographic orientations ($\theta_0=0^\circ$, $\theta_0=\frac{54.7}{2}^\circ$, and $\theta_0=65^\circ$). Each orientation exhibits unique deformation responses, with mesh adaptation reflecting evolving strain distribution at each strain level.

	\end{appendices}

\end{document}